\documentclass[twocolumn,reprint,aps,prb,superscriptaddress]{revtex4-1}

\usepackage{graphicx}

\begin{document}

\title{Optical conductivity in multiferroic GaV$_4$S$_8$ and 
GeV$_4$S$_8$: Phonons and electronic transitions}

\author{S.~Reschke}
 \affiliation{Experimental Physics V, Center for Electronic
Correlations and Magnetism, Institute of Physics, University of Augsburg, 
86135 Augsburg, Germany}

\author{F.~Mayr}
% \email[Corresponding author: ]{Franz.Mayr@Physik.Uni-Augsburg.de}
 \affiliation{Experimental Physics V, Center for Electronic
Correlations and Magnetism, Institute of Physics, University of Augsburg, 
86135 Augsburg, Germany}

\author{Zhe~Wang}
\altaffiliation[Present address: ]{Helmholtz-Zentrum Dresden-Rossendorf, 
Bautzner Landstraße 400, 01328 Dresden, Germany}
 \affiliation{Experimental Physics V, Center for Electronic
Correlations and Magnetism, Institute of Physics, University of Augsburg, 
86135 Augsburg, Germany}

\author{P.~Lunkenheimer}
 \affiliation{Experimental Physics V, Center for Electronic
Correlations and Magnetism, Institute of Physics, University of Augsburg, 
86135 Augsburg, Germany}

\author{W.~Li}
 \affiliation{1. Physikalisches Institut, Universit\"{a}t Stuttgart, 70550 Stuttgart, Germany}

\author{D.~Szaller}
\affiliation{Department of Physics, Budapest University of Technology and 
 Economics and MTA-BME Lend\"{u}let Magneto-optical Spectroscopy Research 
 Group, 1111 Budapest, Hungary}

\author{S.~Bord\'{a}cs} 
\affiliation{Department of Physics, Budapest University of Technology and 
 Economics and MTA-BME Lend\"{u}let Magneto-optical Spectroscopy Research 
 Group, 1111 Budapest, Hungary} 

\author{I.~K\'{e}zsm\'{a}rki}
 \affiliation{Experimental Physics V, Center for Electronic
Correlations and Magnetism, Institute of Physics, University of Augsburg, 
86135 Augsburg, Germany}
 \affiliation{Department of Physics, Budapest University of Technology and 
 Economics and MTA-BME Lend\"{u}let Magneto-optical Spectroscopy Research 
 Group, 1111 Budapest, Hungary}

\author{V.~Tsurkan}
 \affiliation{Experimental Physics V, Center for Electronic
Correlations and Magnetism, Institute of Physics, University of Augsburg, 
86135 Augsburg, Germany}
 \affiliation{Institute of Applied Physics, Academy of Sciences of Moldova, 
 MD-2028 Chisinau, Republic of Moldova}
 
 \author{A.~Loidl}
  \affiliation{Experimental Physics V, Center for Electronic
 Correlations and Magnetism, Institute of Physics, University of Augsburg, 
 86135 Augsburg, Germany}

\date{\today}

\begin{abstract}
We report on optical spectroscopy on the lacunar spinels GaV$_4$S$_8$ and GeV$_4$S$_8$ in the spectral range from 100 to 23\,000\,cm$^{-1}$ and for temperatures from 5 to 300\,K. These multiferroic spinel systems reveal Jahn-Teller driven ferroelectricity and complex magnetic order at low temperatures. We study the infrared-active phonon modes and the low-lying electronic excitations in the cubic high-temperature phase, as well as in the orbitally and in the magnetically ordered low-temperature phases. We compare the phonon modes in these two compounds, which undergo different symmetry-lowering Jahn-Teller transitions into ferroelectric and orbitally ordered phases, and exhibit different magnetic ground states. We follow the splitting of the phonon modes at the structural phase transition and detect additional splittings at the onset of antiferromagnetic order in GeV$_4$S$_8$. We observe electronic transitions within the $d$-derived bands of the V$_4$ clusters and document a significant influence of the structural and magnetic phase transitions on the narrow electronic band gaps.
\end{abstract}

\maketitle

\section{Introduction}

The cluster compounds studied belong to the family of lacunar spinels \textit{A}V$_4$S$_8$, with $A = \mathrm{Ga}$ or Ge, consisting of weakly linked cubane $(\mathrm{V}_4\mathrm{S}_4)^{n+}$ and tetrahedral $(A\mathrm{S}_4)^{n-}$ clusters, arranged in a fcc structure with $F\overline{4}3m$ cubic non-centrosymmetric symmetry at high-temperatures. This structure is derived from a cubic normal spinel \textit{A}V$_2$S$_4$ by removing every second \textit{A}-site ion.  The V$_4$ units form stable, strongly bonded molecules with a unique and collective electron distribution and well-defined spin. In the gallium compound with $n = 5$, vanadium exhibits an average valence of 3.25 and the V$_4$ molecule is characterized by 7 electrons. In the germanium lacunar spinel with $n = 4$, the four V$^{3+}$ ions constitute a molecule with 8 electrons.\cite{mueller:2006, pocha:2000,bichler:2008,chudo:2006,widmann:2016a,widmann:2016b}  In a molecular orbital scheme of V$_4$ clusters, the highest occupied cluster orbitals are triply degenerate with one or with two unpaired electrons in case of the Ga and Ge compounds, respectively. Hence, each V$_4$ cluster carries a spin $S = 1/2$ for the Ga and $S = 1$ for the Ge compound. Due to partial occupation of the triply degenerate electronic levels, both compounds are Jahn-Teller (JT) active. However, the different electronic configurations of the two compounds result in specific JT distortions and consequently in different crystal symmetries of the orbitally ordered phases: In GaV$_4$S$_8$ the V$_4$ tetrahedra are elongated along one of the cubic body diagonals, resulting in a rhombohedral structure with $R3m$ symmetry,\cite{mueller:2006,pocha:2000} while in GeV$_4$S$_8$ the vanadium tetrahedra distort with one long and one short V--V bond on adjacent sites, into an orthorhombic $Imm2$ phase.\cite{mueller:2006,bichler:2008}  In both compounds, the orbitally ordered phases are ferroelectric.\cite{widmann:2016a, widmann:2016b, singh:2014, ruff:2015}  

In the cubic paramagnetic high-temperature phase, both lacunar spinels exhibit moderate antiferromagnetic (AFM) exchange interactions, indicating stable magnetic moments below room temperature. However, in the orbitally ordered phase GaV$_4$S$_8$ exhibits ferromagnetic exchange and subsequently at $T = 12.7$\,K, undergoes a phase transition into a cycloidal phase with ferroelectric excess polarization.\cite{widmann:2016a,ruff:2015} In addition, at low temperatures and moderate magnetic fields the Ga compound hosts a N\'eel-type skyrmion lattice, before a fully spin-polarized state is reached at slightly higher magnetic fields.\cite{kezsmarki:2015}  In clear distinction, in the orbitally ordered phase of GeV$_4$S$_8$, the magnetic exchange remains unaltered and the Ge compound becomes antiferromagnetic below 14.6\,K.\cite{widmann:2016b}
In literature, there exist no reports on further symmetry lowering at the magnetic transition in GaV$_4$S$_8$. AFM order in GeV$_4$S$_8$ seems to be commensurate and has been associated with the space group $Pmn2_1$, which is also compatible with ferroelectricity.\cite{mueller:2006,singh:2014} However, there also exist contradictory reports in literature.\cite{chudo:2006} 

Despite a plethora of interesting phenomena emerging in these compounds, not much is known about phonon properties and about low-lying electronic and orbital transitions in lacunar spinels. 
The phonon modes of GaV$_4$S$_8$ have been studied by Hlinka \textit{et al.}\cite{hlinka:2016} by Raman and infrared spectroscopy. These authors interpreted the phonon spectra at selected temperatures with the aid of \textit{ab-initio} calculations. By combining group-theory analysis and first-principle calculations of electron-phonon coupling constants, Xu and Xiang\cite{xu:2015}  presented a case study on orbital-order driven ferroelectricity using GaV$_4$S$_8$ as a prototypical example. The polar relaxational dynamics at this orbital-order driven ferroelectric transition was studied very recently by THz and broadband dielectric spectroscopy.\cite{wang:2015}  The ferroelectric transition was characterized as order-disorder type, but having first order character. Band structure calculations and resulting magnetic exchange interactions for GaV$_4$S$_8$ have also been published by Zhang \textit{et al.}\cite{zhang:2017} Combined first-principle calculations and an experimental study of the phonon modes in GeV$_4$S$_8$ were performed by Cannuccia \textit{et al.} \cite{cannuccia:2017} In this work, based on the size of the electronic band gap and on phonon eigenfrequencies, the authors speculated that the room temperature space group in the germanium compound may not be $F\overline{4}3m$, but rather $I\overline{4}m2$ and that this space group probably results from dynamic JT distortions, which are present already far above the structural phase transition. Recent THz studies were reported by Warren \textit{et al.}\cite{warren:2017} on polycrystalline and by Reschke \textit{et al.} \cite{reschke:2017b} on single crystalline GeV$_4$S$_8$. 

In the present manuscript, we provide a detailed investigation of the infrared (IR) active phonon modes of GaV$_4$S$_8$ and GeV$_4$S$_8$ for temperatures ranging from 5 to 300\,K. Specifically, we followed the temperature dependencies of eigenfrequencies and damping, focusing on the behavior close to the structural and magnetic phase transitions. In addition, we traced the low-energy electronic transitions below 4\,eV to gain insight into the nature of the electronic band gaps and to check for possible temperature-dependent band shifts and for the influence of phase transitions on the band-gap energies. The influence of magnetic order on optical phonon frequencies\cite{baltensperger:1968, brueesch:1972,sushkov:2005,rudolf:2007b} and anomalous absorption edge shifts in magnetic semiconductors\cite{harbeke:1966, lehmann:1970} has been studied in detail, mainly on spinel compounds already some time ago. In many of these compounds, anomalous absorption-edge shifts were observed when entering the low-temperature magnetic phases due to spin-lattice coupling.

\section{Experimental details}

Optical spectroscopy has been performed on naturally grown $\left(111\right)$ surfaces of well-characterized single crystals. Experimental details concerning sample preparation and characterization of GaV$_4$S$_8$ and GeV$_4$S$_8$ are given in Refs.~\onlinecite{widmann:2016a} and \onlinecite{widmann:2016b}, respectively. The optical reflectivity measurements were carried out using the Bruker Fourier-transform spectrometers IFS~113v and IFS~66v/S, both of them equipped with He-flow cryostats (CryoVac). With the set of mirrors and detectors used for these experiments, we were able to cover a frequency range from 100 to 23\,000\,cm$^{-1}$. In the low frequency region (up to 6\,000\,cm$^{-1}$), a gold mirror was used as reference, while at higher frequencies a silver mirror was utilized. For the evaluation of phonon eigenfrequencies and damping, we directly analyzed the measured reflectivity with a standard Lorentz model for the complex dielectric function with the program RefFIT developed by A.~Kuzmenko,\cite{reffit:1.2.99}  to obtain values of eigenfrequency $\omega_j$, oscillator strength $\Delta\epsilon_j$, and damping parameter $\gamma_j$. The complex dielectric constant has been derived from the reflectivity spectra by means of Kramers-Kronig transformation with a constant extrapolation towards low frequencies and a smooth $\omega^{-1.5}$ high-frequency extrapolation, followed by a $\omega^{-4}$ extrapolation beyond $8 \times 10^5$\,cm$^{-1}$. In analyzing the optical conductivity, we carefully checked the influence of these extrapolation procedures on the results below $2 \times 10^4$\,cm$^{-1}$, relevant for this work.

\section{Results and discussion}

\begin{figure}[htb]
\centering
\includegraphics[width=\linewidth]{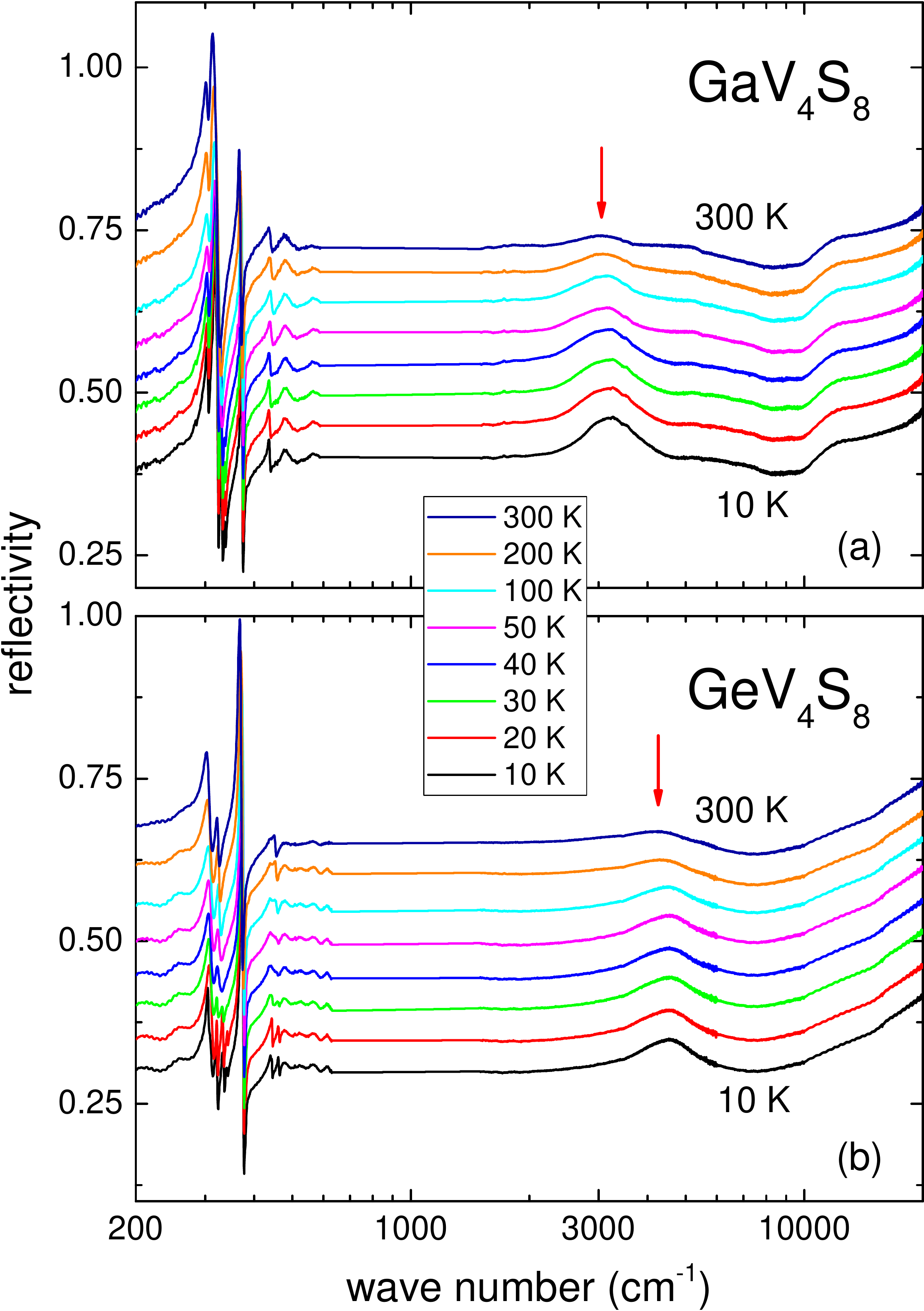}
\caption{Semilogarithmic plot of the frequency dependence of the reflectivity spectra of GaV$_4$S$_8$ (a) and GeV$_4$S$_8$ (b) at a series of temperatures between 10 and 300\,K. The data at 10\,K correspond to the measured reflectivity. The indicated reflectivity scales in (a) and (b) are relevant only for the 10\,K data. For clarity, the reflectivities for the higher temperatures were shifted by a constant offset of 0.05, each. Red arrows in both panels indicate the lowest-energy electronic excitations, i.e. the region of the gap edge.}
\label{fig:1}
\end{figure}

Figure~\ref{fig:1} shows the frequency dependence of the reflectivities measured in GaV$_4$S$_8$ and GeV$_4$S$_8$ for frequencies up to 23\,000\,cm$^{-1}$ for a series of selected temperatures between 10 and 300\,K. Phonons are visible for wavenumbers below 600\,cm$^{-1}$, while the fingerprints of electronic transitions appear between 3\,000\,cm$^{-1}$ and 6\,000\,cm$^{-1}$.  These weak peak-like structures, being the manifestations of the gap edge in the reflectivity, likely correspond to transitions within the $d$ bands of the vanadium V$_4$ molecules. From detailed band-structure calculations it is clear that the electronic density of states near the chemical potential is dominated by bonding states of the V$_4$ molecular clusters and that the insulating band gap in both compounds is determined by electronic transitions within vanadium-derived $d$ bands.\cite{mueller:2006, xu:2015}  $\mathrm{LDA} + U$ calculations yield narrow band gaps of 140\,meV and 200\,meV for the Ga and Ge compound, respectively.\cite{mueller:2006,pocha:2000} Already a first inspection of the reflectivity data makes clear that –- in agreement with the theoretical predictions -- the electronic transitions with vanadium $d$ band character are at higher frequencies in the Ge compound when compared to GaV$_4$S$_8$. In both materials, these transitions become significantly more pronounced on decreasing temperatures. From resistivity measurements, a band gap of 240\,meV has been deduced for the Ga compound,\cite{widmann:2016a} while in GeV$_4$S$_8$ the gap was found to be about 300\,meV.\cite{widmann:2016b} From these measurements it was concluded that, in both compounds, the energy gaps are strongly temperature dependent and that at low temperatures hopping transport dominates the dc resistivity. Concerning phonon excitations, in the cubic high-temperature phase, in Fig.\ref{fig:1} only four phonon modes out of the six IR allowed modes are observed.

\subsection{Phonons}

\begin{figure}[t]
\centering
\includegraphics[width=\linewidth]{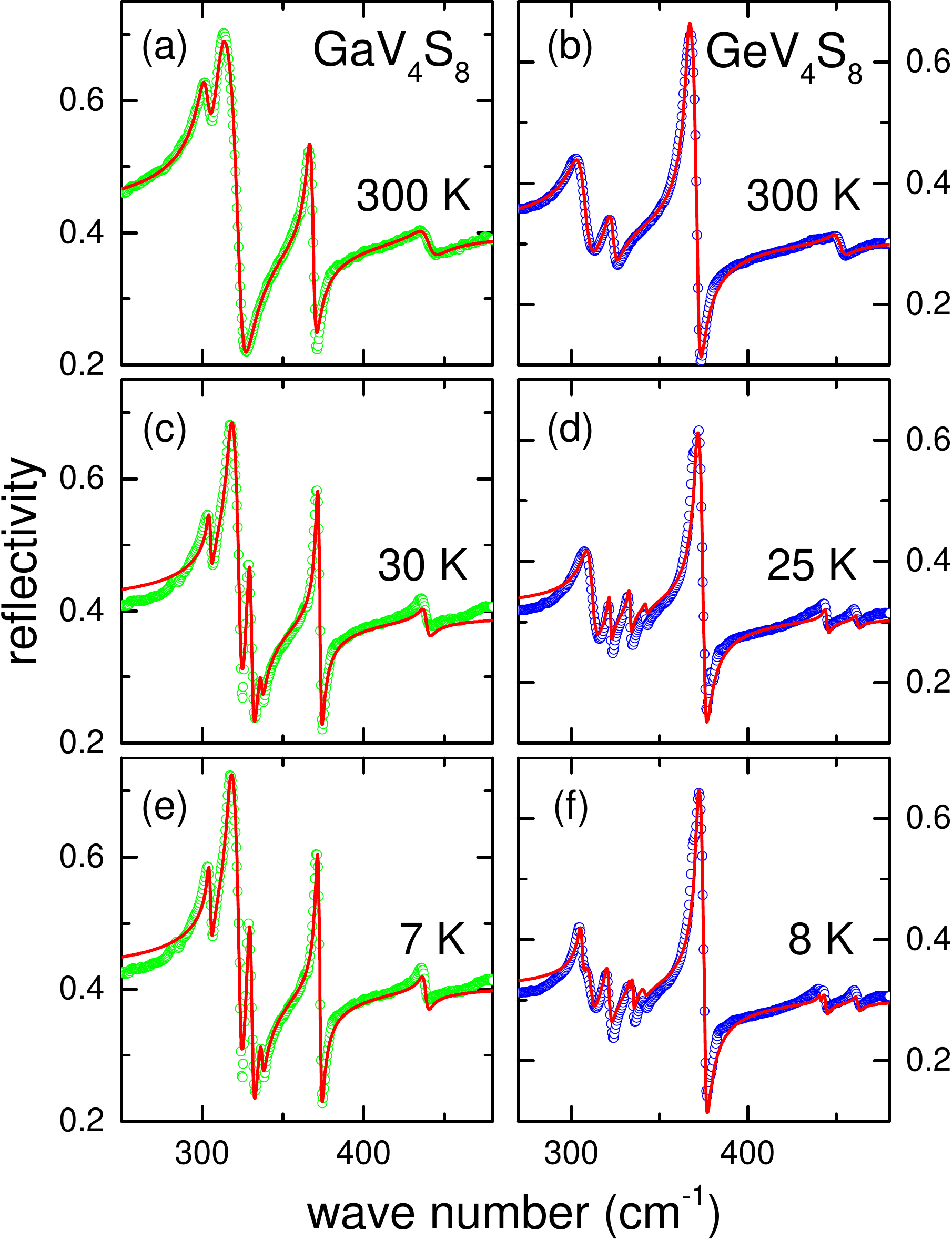}
\caption{Frequency dependence of the phonon spectra of GaV$_4$S$_8$ and GeV$_4$S$_8$ in the FIR range for selected temperatures. Spectra in the cubic high-temperature phases (a, b) are compared to spectra taken in the paramagnetic and orbitally ordered phases (c, d) as well as in the magnetically ordered phases (e, f). The solid lines represent fits to the experimental data as described in the text.}
\label{fig:2}
\end{figure}

The phonon spectra of GaV$_4$S$_8$ and GeV$_4$S$_8$ were measured in the far-infrared (FIR) frequency regime between 5 and 300\,K. Figure~\ref{fig:2} shows a comparative study of the phonon response of GaV$_4$S$_8$ [Figs.~\ref{fig:2}(a), (c), and (e)] and GeV$_4$S$_8$ [Figs.~\ref{fig:2}(b), (d), and (f)], in the cubic high-temperature phases [Figs.~\ref{fig:2}(a) and (b)], in the paramagnetic and orbitally ordered [Figs.~\ref{fig:2}(c) and (d)], as well as in the magnetically and orbitally ordered [Figs.~\ref{fig:2}(e) and (f)] phases. In GaV$_4$S$_8$, four phonon modes can be observed at room temperature. On lowering the temperature, two additional modes appear just below the structural phase transition at $T_\mathrm{JT} = 44$\,K, due to the lowering of the crystal symmetry from cubic to rhombohedral. At the magnetic transition ($T_\mathrm{C} = 12.7$\,K) no further phonon splitting is observed in GaV$_4$S$_8$. In GeV$_4$S$_8$ also four phonon modes appear in the high-temperature cubic phase. Below the symmetry-lowering structural phase transition at $T_\mathrm{JT} = 30.5$\,K, where the crystal symmetry changes from cubic to orthorhombic, three additional modes appear in the phonon spectra of GeV$_4$S$_8$. 

\begin{figure}[t]
\centering
\includegraphics[width=\linewidth]{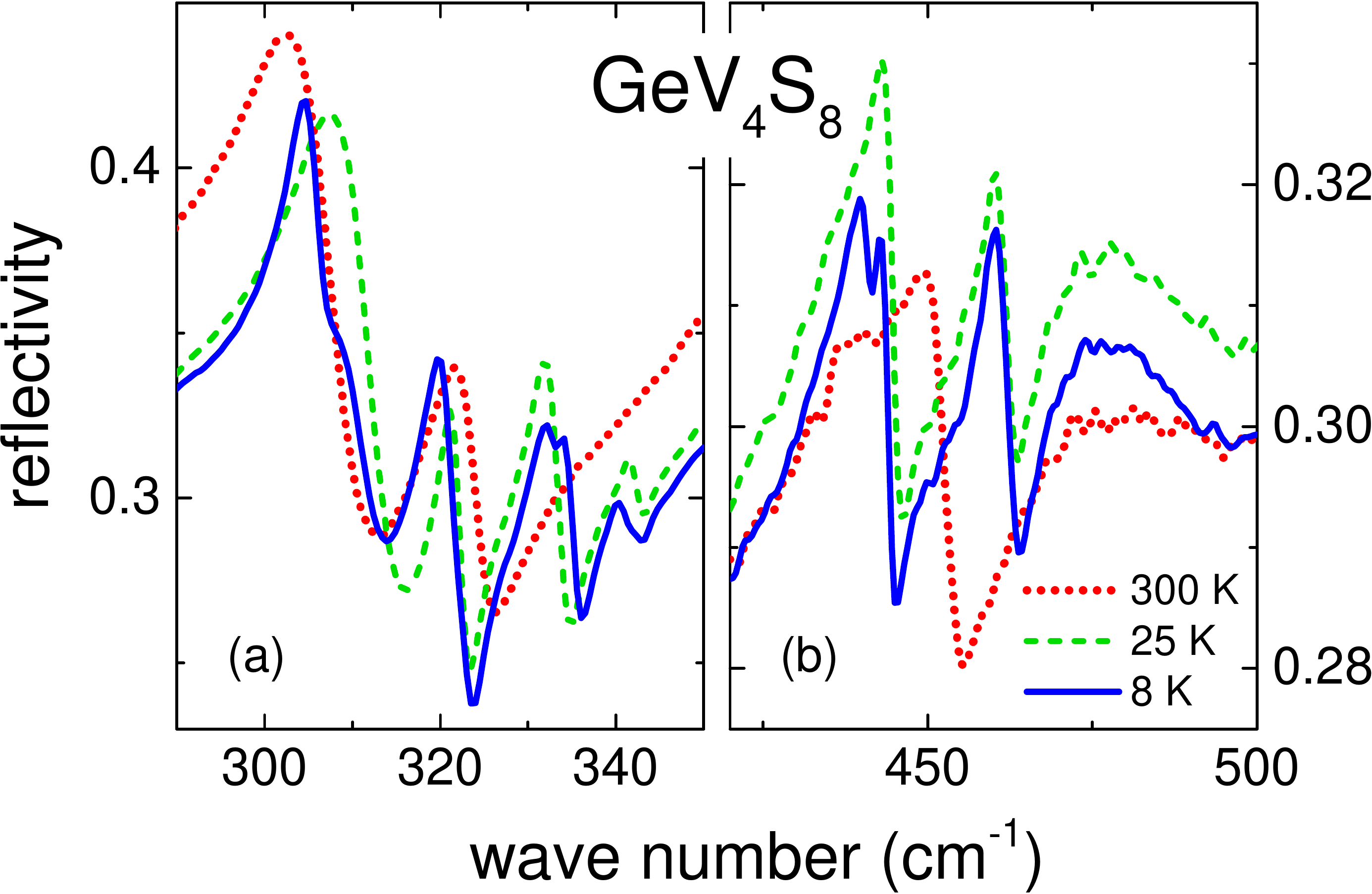}
\caption{Wave number dependencies of the reflectivity in GeV$_4$S$_8$ in the regions from (a) 290 –- 350\,cm$^{-1}$ and (b) 430 –- 500\,cm$^{-1}$, to document the splitting of phonon modes at the JT transition as well as at the onset of AFM order.}
\label{fig:3}
\end{figure}

In contrast to GaV$_4$S$_8$, in GeV$_4$S$_8$ a further small splitting of three phonon modes is detected below the antiferromagnetic phase transition at $T_\mathrm{N} = 14.6$\,K. This splitting is best seen in the lowest frequency mode, just above 300\,cm$^{-1}$, where a shoulder evolves at the high-frequency flank, just at the onset of AFM order. To further document this additional phonon splitting in the Ge compound, Fig.~\ref{fig:3} shows enlarged views of the reflectivity from 300 to 340\,cm$^{-1}$ [Fig.~\ref{fig:3}(a)] as well as from 430 to 500\,cm$^{-1}$ [Fig.~\ref{fig:3}(b)], in the high temperature cubic phase (300\,K), in the paramagnetic and orbitally ordered phase (25\,K) and in the AFM ground state (8\,K). On decreasing temperatures, the phonon mode which appears close to 305\,cm$^{-1}$ in the high-temperature cubic phase [Fig.~\ref{fig:3}(a)], hardens with no indications of splitting at the JT transition and softens again in the magnetically ordered phase. The evolution of a shoulder indicates a splitting of this mode with the onset of AFM order. The mode close to 320\,cm$^{-1}$ [Fig.~\ref{fig:3}(a)] splits at the JT transition into three modes, with one mode undergoing a further splitting at the AFM phase transition. The phonon mode close to 450\,cm$^{-1}$ [Fig.~\ref{fig:3}(b)] undergoes a splitting in two modes at the JT transition, with the lower frequency mode undergoing a further splitting at the onset of AFM order. 

We offer three possibilities to explain the emergence of new modes in the antiferromagnetic phase: i) although the symmetry is orthorhombic already below $T_\mathrm{JT}$, an extra term in the distortion upon magnetic ordering is needed to clearly observe the mode splitting. This interpretation seems to be consistent with x-ray scattering results below the N\'{e}el temperature: In the magnetically ordered phase only a change of the diffraction profiles was reported,\cite{bichler:2008} indicating further distortions without change of symmetry. ii) This mode splitting could be reminiscent of exchange-driven phonon splitting that was found in a number of spinel compounds.\cite{sushkov:2005,rudolf:2007b,kant:2012,bordacs:2009} Finally  iii), antiferromagnetic spin order induces in principle a doubling of the unit cell, and any weak spin-lattice coupling could be responsible for additional modes. Indeed, neutron diffraction revealed a doubling of either the orthorhombic $a$ or $b$ axis below $T_\mathrm{N}$,\cite{mueller:2006} strongly favoring this scenario.  Comparing the IR spectra of the two compounds, at 300\,K the eigenfrequencies of the four experimentally observed modes are very similar, but the intensities exhibit marked differences and document significant alterations of the oscillator strengths and hence, of the ionic plasma frequencies of the modes.

The primitive cell of lacunar spinels contains one formula unit, which gives a total of 39 modes. Group theory predicts $3 A_1 + 3 E + 3 F_1 + 6 F_2$ zone center optical phonon modes in the cubic state, out of which only the $F_2$ modes are IR active.  As clear from Figs.~\ref{fig:2}(a) and (b) only four modes are detected. Reflectivity spectra in the low-frequency range down to 100\,cm$^{-1}$ (not shown) do not reveal additional phonon excitations. Therefore we conclude that these two modes have too low optical weight. This is in accord with results from Hlinka \textit{et al.}\cite{hlinka:2016}, where also only four IR active modes were detected. Two further $F_2$ modes, which were theoretically predicted\cite{hlinka:2016} close to 133 and 203\,cm$^{-1}$ are unobservable. Despite significantly different ionic plasma frequencies, this is also true for the Ge compound.
In THz transmission experiments, at low temperatures an  excitation close to 90\,cm$^{-1}$ was observed for  GeV$_4$S$_8$ (Ref.~\onlinecite{reschke:2017b}), which might be assigned to a calculated $B_2$ phonon mode at 67.4\,cm$^{-1}$,  obtained by density functional theory for the $Imm2$ structure.\cite{cannuccia:2017}

For a detailed analysis of the phonon spectra, we fitted all reflectivity data using the RefFIT fit routine.\cite{reffit:1.2.99} To describe the experimental data, the following sum of Lorentz oscillators was used for the complex dielectric function $\epsilon(\omega)$:
\begin{equation}\label{eq:1}
\epsilon(\omega) = \epsilon_\infty + \sum_{j} \frac{\Delta \epsilon_j 
\omega^{2}_{j}}{\omega^{2}_{j} - \omega^{2} - i \gamma_j \omega}
\end{equation}
Here, $\omega_j$, $\gamma_j$ and $\Delta \epsilon_j$ denote eigenfrequency, damping and the dielectric strength of mode $j$, respectively. $\epsilon_\infty$ corresponds to high-frequency electronic contributions to the dielectric constant. It is useful to define the ionic plasma frequency $\Omega_j$ of mode $j$, which is related to the oscillator strength via
\begin{equation}\label{eq:2}
\Omega^2_j = \Delta \epsilon_j \omega^2_j 
\end{equation}
From the dielectric function $\epsilon(\omega)$, the reflectivity $R(\omega)$ can directly be calculated according to
\begin{equation}\label{eq:3}
R(\omega) = \left| \frac{1 - \sqrt{\epsilon(\omega)}}{1 + 
\sqrt{\epsilon(\omega)}}\right|^2
\end{equation}
Using this approach, we fitted the reflectivity data and the fit results are shown as solid lines in Fig.~\ref{fig:2}. Overall, the fitted spectra  are in satisfactory agreement with the experimental data. Stronger deviations between the fits and experimental data appear in the orbitally and magnetically ordered phases at the lowest wave numbers shown in Fig.~\ref{fig:2}.  These deviations could result from strong relaxational modes\cite{wang:2015} or from low-lying optical phonons\cite{warren:2017} beyond the experimentally accessible spectral range. One could also speculate that low-lying molecular excitations of the vanadium clusters may become relevant at low temperatures.
In systems with strong spin-phonon coupling the line shapes of the phonon modes can become asymmetric.\cite{lee:2004} In these cases Fano-type functions have to be used to obtain better fits. As strong spin-phonon coupling can be expected in the compounds under consideration, we also tried  to improve the fit quality by using Fano line shapes. However, we found no significant improvements of the fits.

\begin{figure}[t]
\centering
\includegraphics[width=\linewidth]{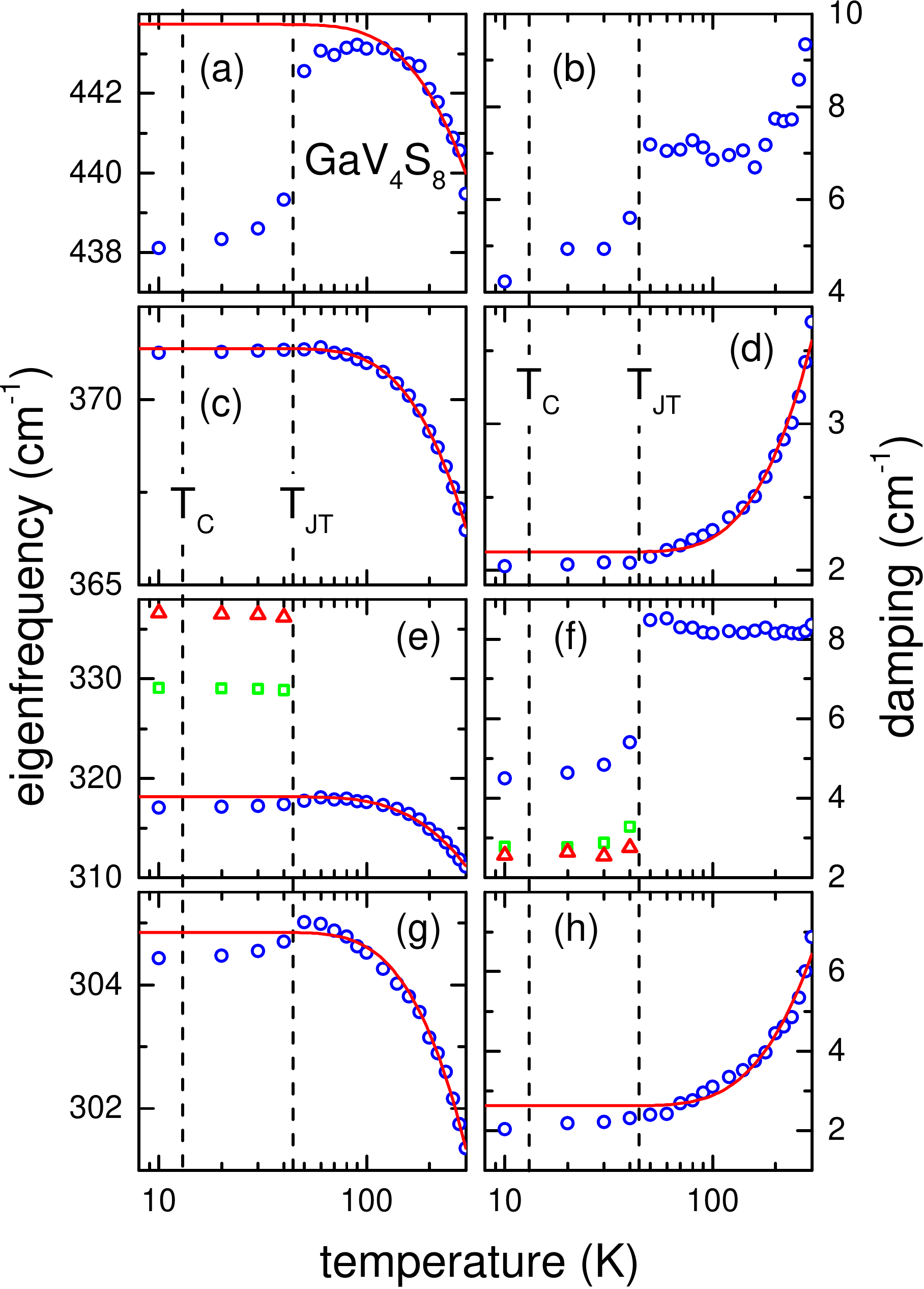}
\caption{Temperature dependence of eigenfrequencies (a, c, e, and g) and damping (b, d, f, and h) of the four phonon modes in GaV$_4$S$_8$ in semilogarithmic representation. The dashed vertical lines indicate the magnetic and the structural phase transition at $T_\mathrm{C} = 12.7$\,K and $T_\mathrm{JT} = 44$\,K, respectively. The solid lines represent fits of the experimental data assuming anharmonic temperature dependences of eigenfrequencies and damping, as described in the text.}
\label{fig:4}
\end{figure}

\begin{figure}[htb]
 \centering
 \includegraphics[width=\linewidth]{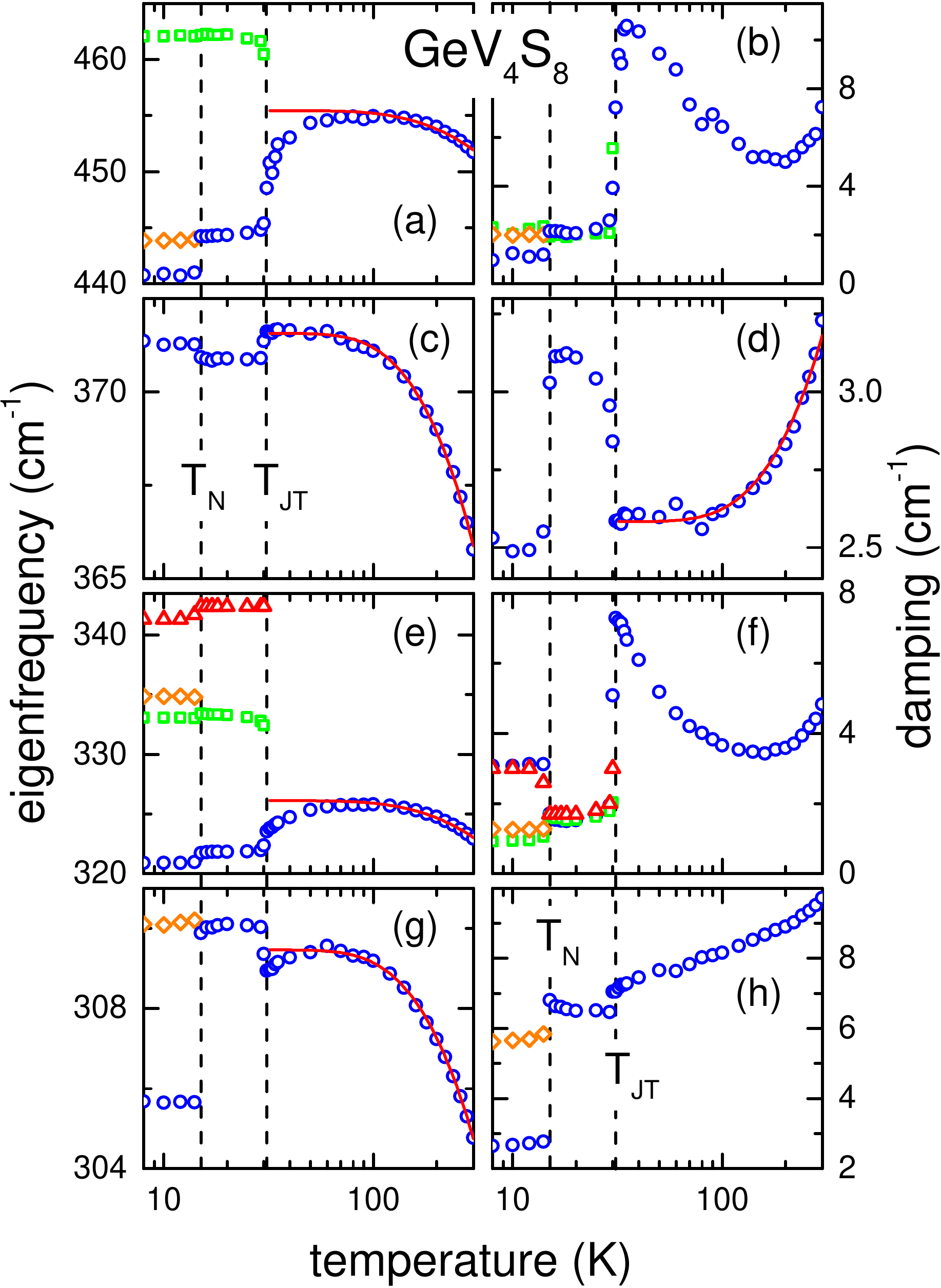}
 \caption{Temperature dependence of eigenfrequencies (a, c, e, and g) and damping (b, d, f, and h)  of the four phonon modes observed in GeV$_4$S$_8$ in semilogarithmic representation. The dashed vertical lines indicate the magnetic and the structural phase transition at $T_\mathrm{N} = 14.6$\,K and $T_\mathrm{JT} = 30.5$\,K, respectively. The solid lines represent fits of the experimental data assuming anharmonic temperature dependences of eigenfrequencies and damping, as described in the text.}
 \label{fig:5}
 \end{figure}

Figures~\ref{fig:4} and \ref{fig:5} show the temperature dependences of eigenfrequency $\omega_j$ and damping $\gamma_j$ of the phonon modes of GaV$_4$S$_8$ and GeV$_4$S$_8$, respectively. For $T > T_\mathrm{JT}$ significant temperature dependences of the eigenfrequencies and damping constants can be observed in both compounds. Neglecting effects, which seem to be strongly correlated with the occurrence of phase transitions, for all phonon modes the eigenfrequencies decrease with increasing temperature, while damping effects rather increase. This continuous decrease of eigenfrequencies and continuous inrease of damping effects on increasing temperatures, is the expected behavior of canonical anharmonic dielectric solids, which mainly result from phonon-phonon interactions.\cite{cowley:1965,klemens:1966,menendez:1984}  In the present work, we do not want to present detailed calculations of phonon frequency shifts and damping constants.  Our main aim is to derive a phenomenological description of the anharmonic behavior and to determine in this way possible shifts in eigenfrequencies or changes in damping, due to the occurrence of orbital order at the JT transition or due to the onset of magnetic order. Following the work of Balkanski \textit{et al.}\cite{balkanski:1983}, the temperature dependence of the eigenfrequencies of an anharmonic solid can be described by the following expression:

\begin{equation}\label{eq:4}
\omega_j(T) = \omega_{0j} \cdot \left( 1 - \frac{c_j}{\exp (\frac{\hbar \Omega}{k_{B} T}) - 1} 
\right)
\end{equation} 

In this simplified anharmonic model the damping increases with increasing temperature as:

\begin{equation}\label{eq:5}
\gamma_j(T) = \gamma_{0j} \cdot \left( 1 + \frac{d_j}{\exp (\frac{\hbar \Omega}{k_{B} T}) - 
1} 
\right)
\end{equation} 

Here $\omega_{0j}$ and $\gamma_{0j}$ denote eigenfrequency and the damping of mode $j$ at $T = 0$\,K. $\hbar \Omega$ corresponds to the energy of the characteristic frequency of the decaying phonon modes. The parameters $c_j$ and $d_j$ are additional parameters, which describe the strengths of the anharmonic contributions for eigenfrequencies and damping for different modes. This ansatz for the temperature dependence of frequency shift and damping is only based on three-phonon processes and has to be viewed as a mere parametrization of anharmonic contributions. It has been shown experimentally that four phonon processes play a minor role only and usually can be neglected.\cite{choi:2003}

For GaV$_4$S$_8$, the temperature dependence of the eigenfrequencies above the structural phase transition can be described by Eq.~(\ref{eq:4}), with an average characteristic frequency $\Omega$ of 243\,cm$^{-1}$ for all four phonon modes, as indicated by the solid lines in Fig.~\ref{fig:4}. Deviations from anharmonic behavior mainly occur near the structural phase transition. This especially is true for the high-frequency phonon mode [Fig.~\ref{fig:4}(a)], where the eigenfrequency softens when approaching the phase transition from high temperatures and strongly decreases below $T_\mathrm{JT}$. For the phonon modes shown in Figs.~\ref{fig:4}(e) and (g), the eigenfrequencies decrease for $T < T_\mathrm{JT}$. Only the phonon mode close to 370\,cm$^{-1}$ is almost unaffected by the structural phase transitions [Fig.~\ref{fig:4}(c)]. Due to the symmetry-lowering phase transition, for $T < T_\mathrm{JT}$ two additional phonons appear at about 329\,cm$^{-1}$ and 336\,cm$^{-1}$ in the orbitally ordered phase [Fig.~\ref{fig:4}(e)]. Below $T_{JT}$, GaV$_4$S$_8$ has a rhombohedral structure with $R3m$ symmetry. According to group theory and using the Wyckoff positions from Ref.~\onlinecite{pocha:2000}, we expect a total of 21 IR active optical modes at the zone center. Only six modes have been identified, documenting that most of the modes have very low optical weight. It is interesting that the eigenfrequencies remain almost unchanged when crossing the magnetic phase transition. In addition, no further phonon modes were detected in the low-temperature magnetic phase with cycloidal spin structure. Concerning the temperature dependence of the phonon damping, the phonon modes close to 304 [Fig.~\ref{fig:4}(h)] and 370\,cm$^{-1}$ [Fig.~\ref{fig:4}(d)] can be approximately described by purely anharmonic effects for all temperatures using Eq.~(\ref{eq:5}). Here we used the same characteristic frequency $\Omega$, which was deduced already for the temperature dependence of the frequency shifts.  Significant damping effects appear for the phonon modes at 330\,cm$^{-1}$ [Fig.~\ref{fig:4}(f)] and 440\,cm$^{-1}$ [Fig.~\ref{fig:4}(b)]. For these modes, the damping is extremely strong in the high-temperature cubic phase and becomes suppressed in the orbitally ordered phase. These phonon modes probably are strongly susceptible to orbital reorientations when approaching the Jahn-Teller transition.
 
As revealed by Fig.~\ref{fig:5}, the temperature dependence of the phonon modes in GeV$_4$S$_8$ behave significantly different compared to the Ga compound. The eigenfrequencies almost follow an anharmonic behavior in the cubic phase, undergo slight softening when approaching the structural phase transitions and show significant jumps or splittings in the orbitally ordered phase ($T < T_\mathrm{JT} = 30.5$\,K). Additional modes also emerge at the magnetic phase transition. Clear indications of symmetry-lowering splittings are found in GeV$_4$S$_8$, upon passing the antiferromagnetic phase transition at 14.6 K. We would like to recall that no splitting of phonon modes was observed in GaV$_4$S$_8$ when passing the magnetic phase transition, where cycloidal spin order is established at low temperatures. In the Ge compound, we observe a total of 7 modes in the orbitally ordered phase, whereas 10 modes appear in the antiferromagnetic phase. The orbitally ordered phase in GeV$_4$S$_8$ is orthorhombic with $Imm2$ symmetry. Using the Wyckoff positions of Ref.~\onlinecite{bichler:2008}, we expect 30 IR active modes at the zone center. From those 30 modes, only a small fraction is observed. The rest seems to have too low optical weight. 

While the splitting of phonon modes or the appearance of new modes at a symmetry lowering structural phase transition follows quite naturally on the basis of symmetry considerations, the splitting of modes at the onset of AFM order is only rarely observed. The influence of magnetic super-exchange interactions on optical phonon frequencies has been studied in detail almost 50 years ago.\cite{baltensperger:1968,brueesch:1972} Later-on, the splitting of phonon modes in antiferromagnets due to spin-phonon coupling has been explained in terms of a spin Jahn-Teller effect\cite{sushkov:2005} and a linear dependence of the phonon splitting on the exchange coupling has been documented for a number of transition-metal oxides.\cite{kant:2012} It demands further modeling and a detailed knowledge of the low-temperature structure in GeV$_4$S$_8$ to arrive at definite conclusions about the origin of the appearance of a number of new phonon modes in the AFM phase.

The damping for the modes close to 450 and 330\,cm$^{-1}$ [Figs.~\ref{fig:5}(b) and (f)] seems to be dominated by strong orbital fluctuations above the phase transition and it becomes suppressed below $T_\mathrm{JT}$. Only the mode close to 370\,cm$^{-1}$ [Fig.~\ref{fig:5}(d)] can roughly be described by anharmonic effects in the cubic high-temperature phase. Here the damping becomes strongly enhanced in the orbitally ordered and paramagnetic phase, but is suppressed again when antiferromagnetic order is established.

Without detailed microscopic ab-initio modeling it seems almost impossible to explain jumps in eigenfrequencies and damping. One should have in mind that lacunar spinels belong to the rare class of materials, in which the JT transition induces ferroelectricity, which is rather unusual and has been explained in a case study on GaV$_4$S$_8$ using group theory analysis in Ref.~\onlinecite{xu:2015}. In this study JT distortions were predicted in a model including electron-phonon coupling. However, even in this case no definite predictions about changes of phonon eigenfrequencies at the Jahn-Teller transition can be made. To do so, the symmetry of low and high temperature phases have to be considered as well as the vibrational pattern and eigenvectors of these specific modes. Things become even more difficult as the JT phase transitions in both compounds under consideration are strongly of first order.\cite{widmann:2016a,widmann:2016b} Overall, the experimentally observed behavior indicates that in both compounds the high-frequency phonon mode and the mode close to 330\,cm$^{-1}$ are strongly influenced by the JT transition, despite the fact that in the two compounds orbital order induces very different symmetries in the orbitally ordered phases. This is true for both, eigenfrequencies and damping.

Changes of eigenfrequencies and damping on passing the onset of magnetic order are minor in the Ga compound. In the Ge compound we identified additional splitting of the phonon modes when passing the AFM ordering temperature. Based on this fact, we conclude that spin-phonon coupling will be strong. In principle, it should be possible to analyze phonon splittings or phonon shifts induced by magnetic exchange interactions. It has been shown by Baltensperger and Helman\cite{baltensperger:1968} and by Br\"{u}esch and D'Ambrogio\cite{brueesch:1972} that magnetic exchange interactions influence the force constants and hence the frequencies of lattice vibrations. The frequency shifts of optical phonon modes can be positive or negative, depending on the dominant exchange interactions.\cite{wakamura:1988} However, such an analysis is far beyond the scope of this work. 

 \begin{table*}[tb]
  \caption{Eigenfrequencies of the transverse ($\omega_\mathrm{TO}$) and longitudinal optical ($\omega_\mathrm{LO}$) IR active phonon modes in GaV$_4$S$_8$ and GeV$_4$S$_8$ as experimentally observed at 100\,K. In addition, the high-frequency dielectric constant, $\epsilon_\infty$, and the oscillator strengths, $\Delta \epsilon_j$, of all phonon modes $j$ are given, as well as the optical strength over all experimentally observed IR-active modes.}
  \label{tab:1}
  \begin{ruledtabular}
  \begin{tabular}{cccccc}
  \multicolumn{6}{c}{IR active modes at 100\,K} \\
  \multicolumn{3}{c}{GaV$_4$S$_8$} &  \multicolumn{3}{c}{GeV$_4$S$_8$}  \\ 
  \multicolumn{3}{c}{$\epsilon_\infty = 19.2$} &  
  \multicolumn{3}{c}{$\epsilon_\infty = 12.6$}   \\ 
  $\omega_\mathrm{TO}$ (cm$^{-1}$) & $\omega_\mathrm{LO}$ (cm$^{-1}$) & $\Delta \epsilon_j$ & $\omega_\mathrm{TO}$ (cm$^{-1}$) & $\omega_\mathrm{LO}$ (cm$^{-1}$) & $\Delta \epsilon_j$ \\ 
  \hline
  304.5 & 305.4 & 0.12 & 309.2 & 311.8 & 0.22 \\ 
  317.6 & 327.1 & 1.17 & 325.8 & 326.5 & 0.06 \\ 
  371.0 & 372.7 & 0.18 & 371.1 & 375.6 & 0.31 \\ 
  443.1 & 443.7 & 0.05 & 454.9 & 455.3 & 0.02 \\ 
  %
%    & & $\sum \Delta \epsilon_j = 1.52$ &  &  & $\sum \Delta \epsilon_j = 0.61$ \\ 
     \multicolumn{3}{r}{$\sum \Delta \epsilon_j = 1.52$} &  \multicolumn{3}{r}{$\sum \Delta \epsilon_j = 0.61$} \\ 
  \end{tabular}
  \end{ruledtabular}
  \end{table*}
 
Table~\ref{tab:1} summarizes the main experimental results of this phonon investigation in the high-temperature cubic phase at 100\,K. The frequencies of the longitudinal optical (LO) modes can be calculated via the ionic plasma frequency, $\Omega_j$, and oscillator strength $\Delta \epsilon_j$. In this case the transverse optical (TO) mode corresponds to the eigenfrequency $\omega_j$. The results in GaV$_4$S$_8$ are in relatively good agreement with those of Ref.~\onlinecite{hlinka:2016} deduced at 80\,K. 
Optical phonon frequencies for GeV$_4$S$_8$ at room temperature, as well as in the ferroelectric phase, were also reported by Cannuccia \textit{et al.}\cite{cannuccia:2017} At room temperature these authors found transverse optical phonon modes at 305, 323, 366 and 450\,cm$^{-1}$, not too far from the values reported in this work (see Tab.~\ref{tab:1} and Fig.~\ref{fig:4}). In the orbitally ordered and ferroelectric phase, they observed 6 modes compared to 7 modes, as documented in Fig.~\ref{fig:4} of the present work. Obviously, they missed the splitting of the mode close to 450\,cm$^{-1}$, as shown in Fig.~\ref{fig:3}(b). We would like to recall that from the observed eigenfrequencies in the Ge compound, Cannuccia \textit{et al.} argued about a possible $I\overline{4}m2$ structure which is stabilized by strong dynamic JT distortions already far above the structural phase transition. Reference~\onlinecite{cannuccia:2017} did not report on further splittings when passing the magnetic phase transition.

As became already clear from a first inspection of Fig.~\ref{fig:2}, the eigenfrequencies of the Ga and the Ge compound are very similar, which certainly stems from the fact that the two elements are neighbors in the periodic table of elements with very similar masses. However, the oscillator strengths are significantly different. The difference is enormous for the mode close to 320\,cm$^{-1}$, being almost by a factor of 20 stronger for the Ga compound, signaling that very different effective charges are involved in this lattice vibration. Hence, the overall dipolar strength in the Ge compound is by more than a factor of two lower. This strongly reduced phononic dipolar strength signals very different ionicity of the two compounds, with GaV$_4$S$_8$ being the one with higher ionicity, while GeV$_4$S$_8$ obviously is dominated by significantly stronger covalent bonds. It is also interesting that the high-frequency dielectric constant ($\epsilon_\infty$) of the Ga compounds is significantly higher in comparison with the Ge compound. Generally speaking, smaller high-frequency dielectric constants usually correspond to larger band gaps and this indeed seems to be the case for these two compounds. The values of the high-frequency dielectric constants from this optical study, plus the oscillator strengths of the IR active modes (20.7 and 13.2 for the Ga and Ge compound, respectively) can be compared with results from THz spectroscopy: At THz frequencies, values of 15 and 9.2 were determined for the dielectric constants of GaV$_4$S$_8$ (Ref.~\onlinecite{wang:2015}) and GeV$_4$S$_8$ (Ref.~\onlinecite{warren:2017}), respectively.  While the absolute values differ by about 30\,\%, the ratios of the dielectric constants nicely agree.

 \subsection{Electronic transitions}
 
 To get a more detailed understanding of the low-lying electronic transitions and specifically, to understand the temperature dependence of the electrical resistivity, which exhibits no simple Arrhenius-like behavior over an extended temperature range,\cite{widmann:2016a, widmann:2016b} we studied the optical conductivity of both compounds in the wave-number regime where the conductivity is dominated by electronic transitions within the $d$ bands of the vanadium V$_4$ molecules. At this point, it is rather unclear if the vanadium $d$ electrons are delocalized (metallic) within the vanadium clusters and only can hop from cluster to cluster, resulting in a semiconducting behavior of the electrical resistance. Note that in the Ga compound the average valence of the vanadium is 3.25, while it is 4 in GeV$_4$S$_8$, resulting in V$_4$ molecules with 7, respectively 8 electrons. It is interesting to mention that some type of charge ordering has been reported for the Ge compound at low temperatures.\cite{mueller:2006}
 
 \begin{figure}[t]
  \centering
  \includegraphics[width=\linewidth]{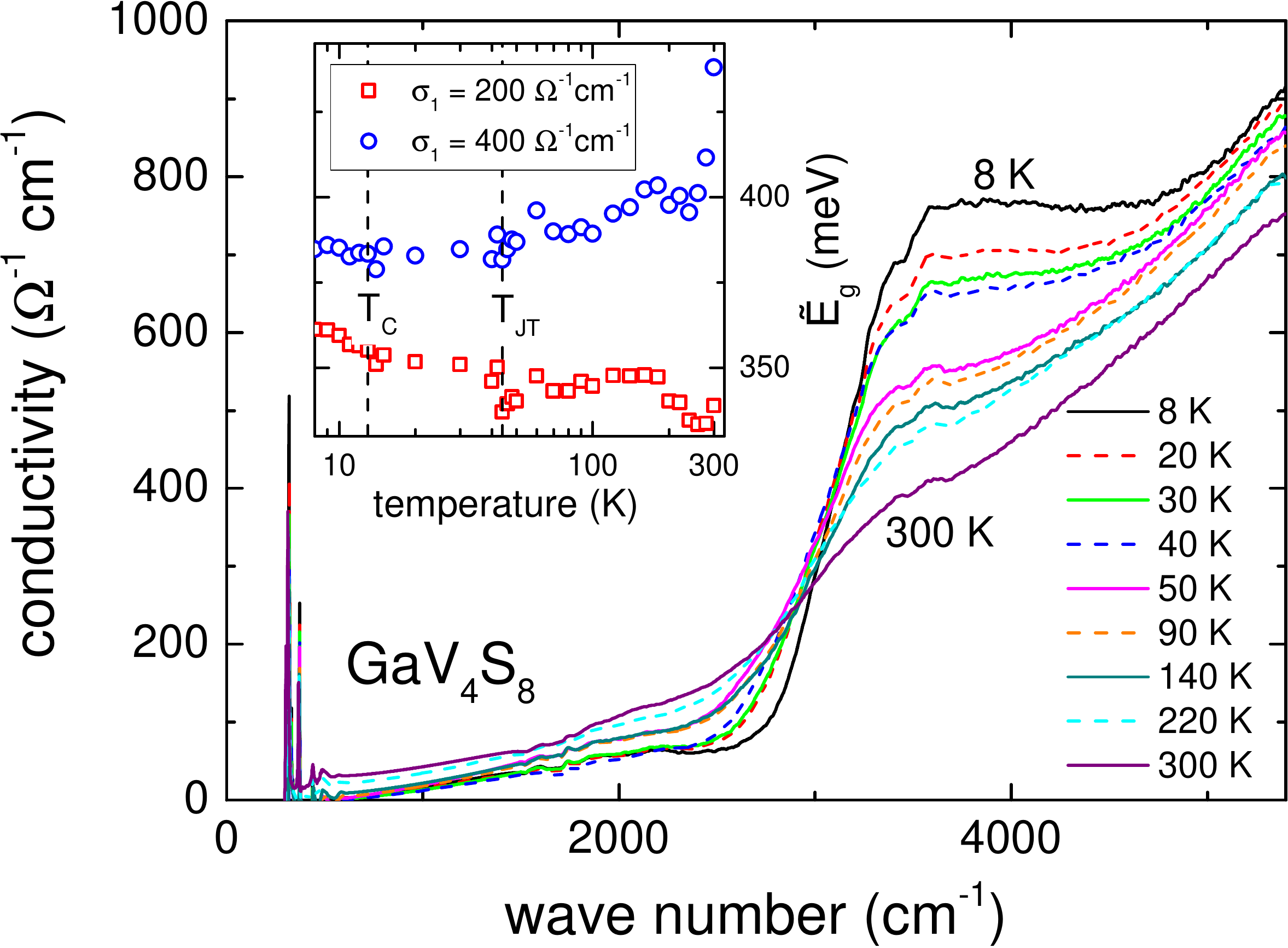}
  \caption{Temperature dependence of the optical conductivity of GaV$_4$S$_8$ for wave numbers up to 5\,200\,cm$^{-1}$ for selected temperatures between 8 and 300\,K. The inset shows the temperature dependence of the frequency (in meV) at constant conductivity values of 200 and 400\,$\Omega^{-1} \mathrm{cm}^{-1}$. These values provide an estimate of an apparent energy gap $\tilde{E}_\mathrm{g}$ (see text). Vertical dashed lines in the inset indicate structural and magnetic ordering temperatures.}
  \label{fig:6}
  \end{figure} 
 
 Figure~\ref{fig:6} shows the frequency dependence of the optical conductivity of GaV$_4$S$_8$ up to 5\,000\,cm$^{-1}$ for a series of temperatures between 8\,K and room temperature. Already at first sight, we can identify a strong smearing out of the band edge close to 3\,000\,cm$^{-1}$ with increasing temperature. While at 8\,K the electronic transition seems to be well defined, resulting in a step-like band edge close to 3\,000\,cm$^{-1}$, at 300\,K the band edge shows an unusual soft-edge behavior. Similar soft-edge behavior was identified in the optical conductivity of the prototypical correlated electron system, the antiferromagnetic insulator V$_2$O$_3$, where the conductivity just above the band gap raises with a conductivity exponent of $3/2$.\cite{thomas:1994}  However, we think that in the case of lacunar spinels, with a Jahn-Teller transition inducing orbital order at low temperatures, this soft-edge behavior probably emerges from strong orbital fluctuations dominating the cubic high-temperature phase. This we will outline and discuss later in more detail. Interestingly, an isosbestic point\cite{greger:2013}   with temperature-independent conductivity appears close to 3\,000\,cm$^{-1}$. Isosbestic points often occur in electronically correlated materials and isosbestic features have been observed and analyzed in THz spectra of iron-based superconductors\cite{wang:2014,wang:2016} and of GeV$_4$S$_8$.\cite{warren:2017} 
 
 As outlined earlier, ferromagnetic and semiconducting spinel compounds very often show an unusual shift of the absorption edge as function of temperature.\cite{harbeke:1966,lehmann:1970} To document this anomalous behavior of the band edge, the temperature dependence of the band gap has been deduced at a constant conductivity value.\cite{harbeke:1966} We were unable to fit unambiguously our $\sigma(\omega)$ curves with a specific frequency dependence for a direct or indirect gap. In order to avoid possible errors by a wrong extrapolation of the wave number dependence of the conductivity, here we also follow the procedure of Ref.~\onlinecite{harbeke:1966} to get a rough estimate of the band-edge. Therefore, the apparent energy gap $\tilde{E}_\mathrm{g}$ was extracted  by taking the intersection of $\sigma(\omega)$ with a horizontal line at a constant conductivity value.
 In the inset of Fig.~\ref{fig:6}, we show the temperature dependent shift of this absorption edge at constant conductivities of 200 and 400\,$\Omega^{-1} \mathrm{cm}^{-1}$, i.e. just below and above the isosbestic point.  Due to significant smearing effects of the band edge, $\tilde{E}_\mathrm{g}$ below the isosbestic point decreases, while above it increases with increasing temperature. No significant anomalies can be detected, neither at the structural nor at the magnetic phase transition. The apparent energy gap as determined from these measurements at room temperature and at lower conductivities is close to 350\,meV ($\approx 2\,800$\,cm$^{-1}$), significantly higher than the value of 240\,meV as determined from the temperature dependence of the electrical resistance in the temperature range 70 to 150\,K.\cite{widmann:2016a}
 
 \begin{figure}[b]
 \centering
 \includegraphics[width=\linewidth]{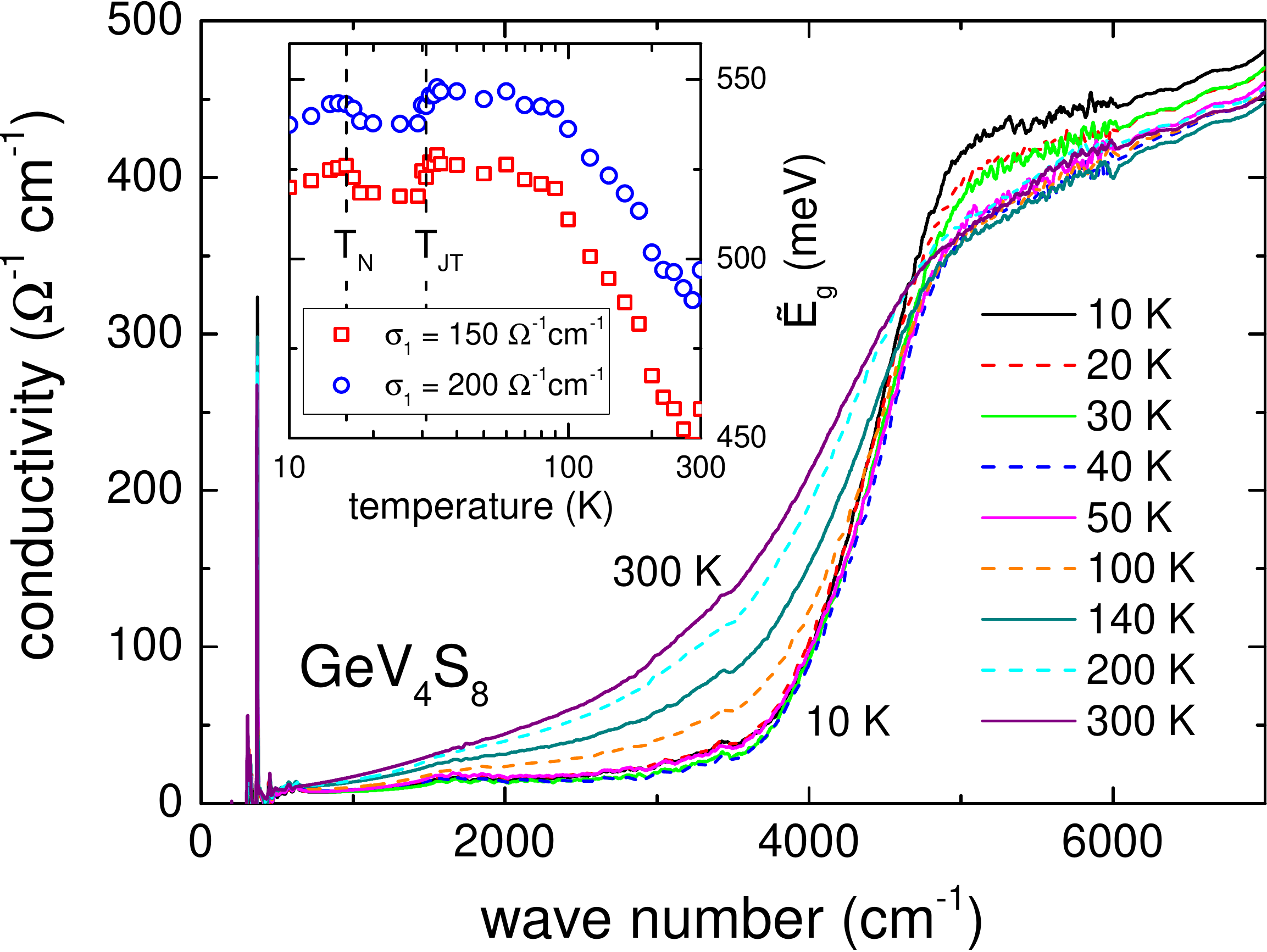}
 \caption{Temperature dependence of the optical conductivity of GeV$_4$S$_8$ for wave numbers up to 7\,000\,cm$^{-1}$ for some selected temperatures between 10 and 300\,K. The inset shows the temperature dependence of the frequency (in meV) at constant values of 150 and 200\,$\Omega^{-1} \mathrm{cm}^{-1}$. These values provide an estimate of an apparent energy gap $\tilde{E}_\mathrm{g}$. Vertical dashed lines in the inset indicate structural and magnetic ordering temperatures.}
 \label{fig:7}
 \end{figure} 

The frequency and temperature dependence of the optical conductivity for GeV$_4$S$_8$ is shown in Fig.~\ref{fig:7}. We find a similar evolution of a soft-edge behavior of the band gap as function of temperature: A well-defined step-like band edge in the magnetically and orbitally ordered phase at 10\,K evolves into a soft-edge behavior in the high-temperature cubic phase at room temperature. The latter, however, is not as drastic as observed in the Ga compound and the average band gap now has shifted to approximately 4\,000\,cm$^{-1}$. Again, an isosbestic point appears in the frequency- and temperature-dependent conductivity, but now at somewhat higher wave numbers close to 5\,000\,cm$^{-1}$. In the Ge compound this isosbestic point also signals a strong shift of optical weight from lower to higher wave numbers for decreasing temperatures. We followed the same procedure as outlined above to determine an apparent temperature-dependent band gap. Here both values of constant conductivity are chosen below the isosbestic point. The results are shown in the inset of Fig.~\ref{fig:7}. In the case of GeV$_4$S$_8$, we find a blue shift of the apparent band edge with decreasing temperature, which amounts almost 10\,\% and certainly cannot be explained by thermal expansion. This blue shift saturates at low temperatures below the Jahn-Teller transition. Quite astonishingly, in the temperature dependence of the band edge we find clear signatures of the structural and the magnetic phase transitions: The apparent band gap slightly decreases at the Jahn-Teller transition and again increases roughly by the same amount when entering the antiferromagnetic phase. The apparent band gap as determined at the conductivity of 150\,$\Omega^{-1} \mathrm{cm}^{-1}$ and at room temperature approximately amounts 450\,meV and is significantly larger than the gap values of 300\,meV determined from electrical resistance measurements.\cite{widmann:2016b}

To further elucidate the energy gaps of GaV$_4$S$_8$ and GeV$_4$S$_8$, we analyzed the energy dependence of the optical conductivities in more detail. In canonical semiconductors, strict power-law dependences are expected, with exponents depending on the nature of the band gap, with electronic transitions being direct or indirect and IR allowed or forbidden. However, we think that the tremendous smearing of the band energies results from orbital fluctuations in the cubic high-temperature phases. Hence, it seems more appropriate just to perform linear extrapolations of the optical conductivities to zero to get an estimate of the true band gaps that dominate these materials.

\begin{figure}[b]
\centering
\includegraphics[width=\linewidth]{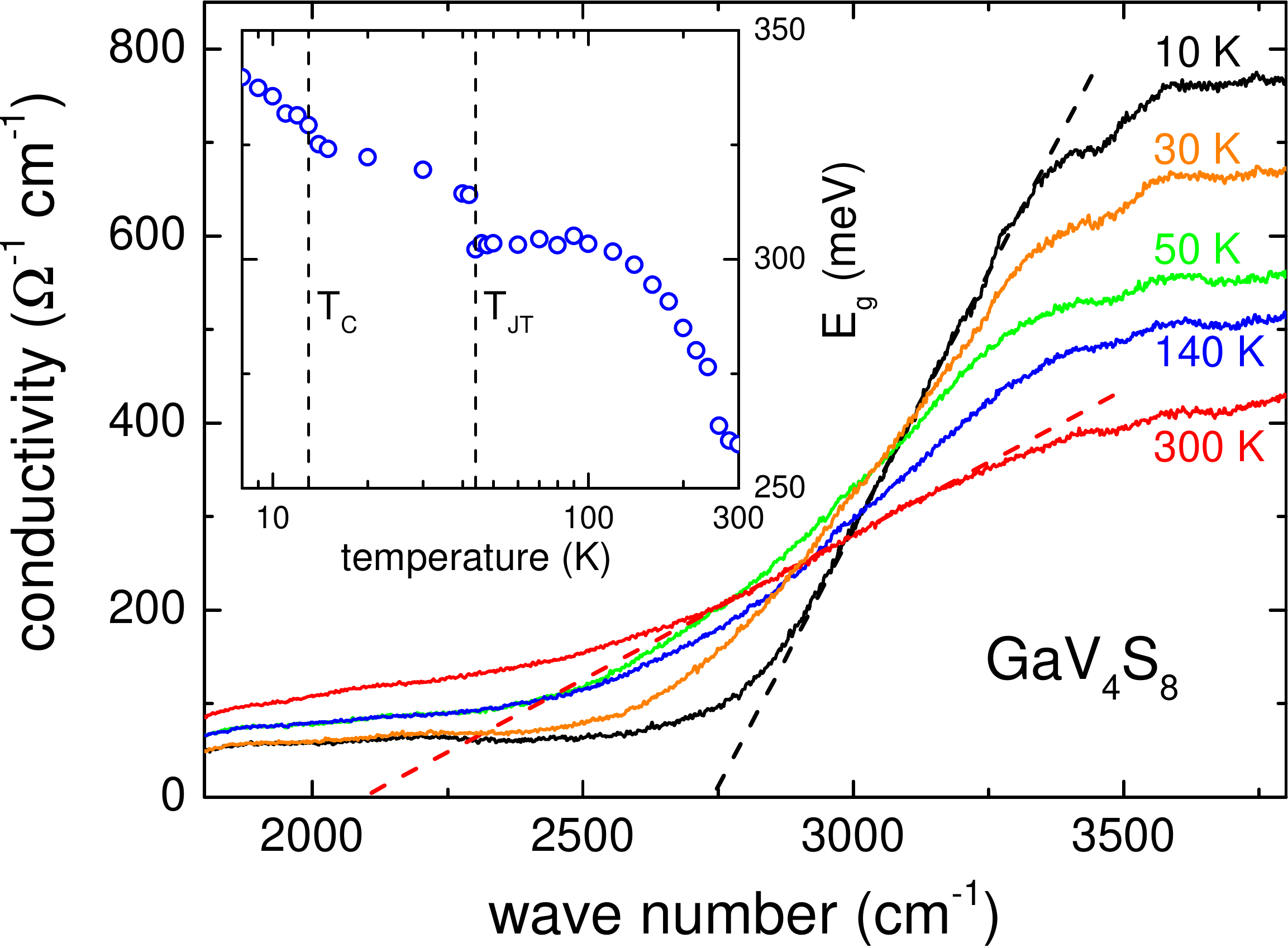}
\caption{Linear extrapolation of the optical conductivity in GaV$_4$S$_8$ to determine the size of the energy gap. Dashed lines indicate the linear extrapolation procedure. The temperature dependence of the gap values is documented in the inset. Vertical dashed lines indicate structural and magnetic ordering temperatures.}
\label{fig:8}
\end{figure} 

In Fig.~\ref{fig:8}, this procedure is documented for the Ga compound for a series of temperatures in the magnetic, the paramagnetic and orbitally ordered, and in the high-temperature cubic phases. For the highest and the lowest temperature, dashed lines indicate the linear extrapolation, which was used to analyze the temperature dependence of the band gap. The obtained values are indicated in the inset of Fig.~\ref{fig:8}. First of all, because of this extrapolation towards zero conductivity, now the band gaps are much closer to the values as determined from the electrical resistance. At room temperature, the value close to 250\,meV corresponds nicely to the values of 240\,meV, determined from the temperature dependence of the resistivity determined at lower temperatures.\cite{widmann:2016a} As in the Ge compound, we find a blue shift of the band gap on decreasing temperatures and it reaches values of approximately 350\,meV at the lowest temperatures. Astonishingly, this linear extrapolation yields clear anomalies in the temperature dependence of the band gap at the structural as well as at the magnetic phase transition. In contrast, no anomalies were detected in the apparent band gaps (inset of Fig.~\ref{fig:6}).

\begin{figure}[t]
\centering
\includegraphics[width=\linewidth]{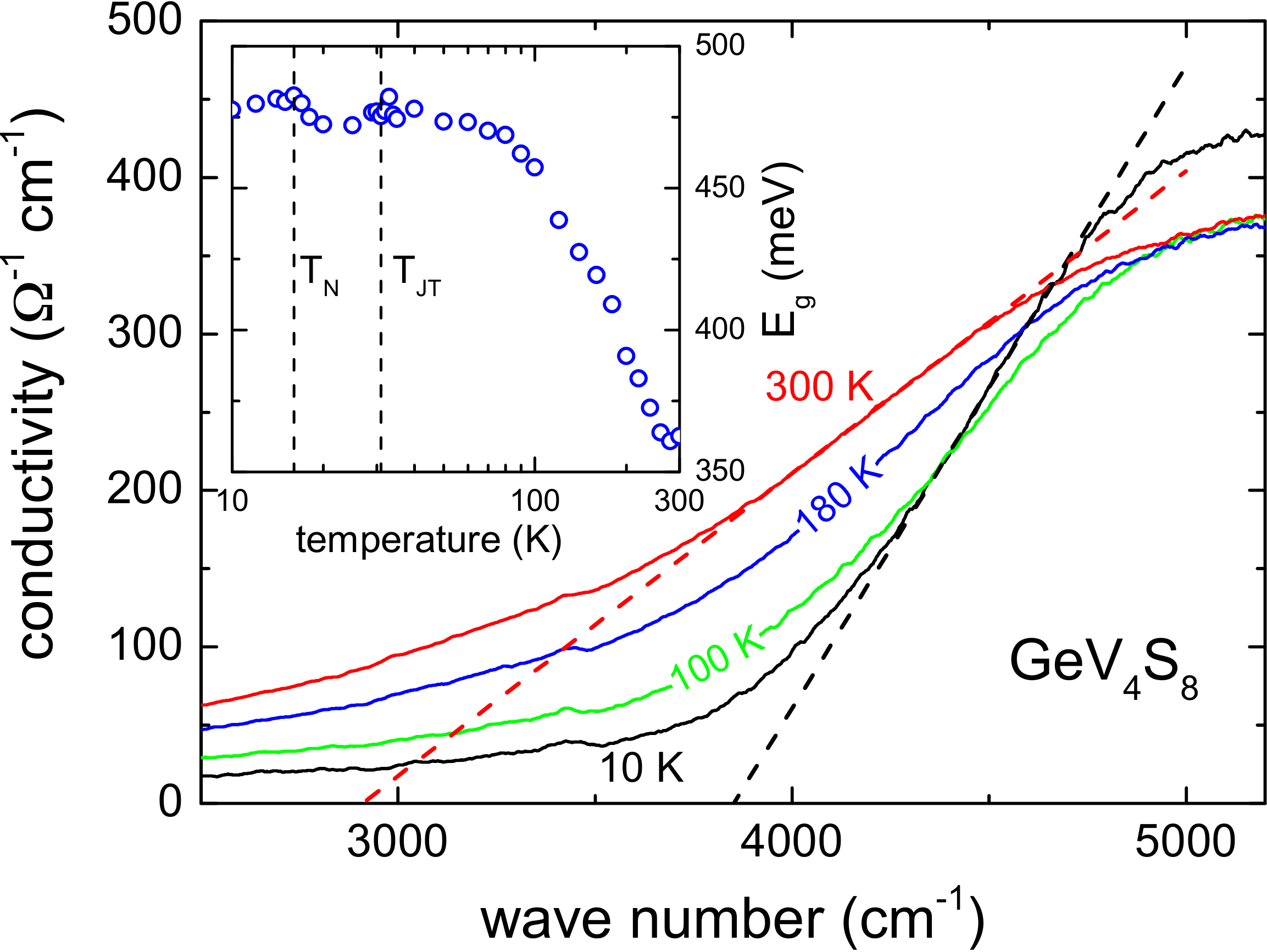}
\caption{Linear extrapolation of the optical conductivity in GeV$_4$S$_8$ to determine the size of the energy gap. Dashed lines indicate the linear extrapolation procedure. The temperature dependence of the gap values is documented in the inset. Vertical dashed lines indicate structural and magnetic ordering temperatures.}
\label{fig:9}
\end{figure} 

A very similar evaluation is shown in Fig.~\ref{fig:9} for the optical conductivities in GeV$_4$S$_8$. Again, the extrapolation procedure to determine the temperature dependence of the band gap is documented by dashed lines for the lowest and highest temperatures.  The results are indicated in the inset of Fig.~\ref{fig:9}. On decreasing temperature, the band gap increases from 350\,meV up to 475\,meV at the lowest temperatures. From electrical resistivity measurements the insulating gap in the Ge compound was of the order of 300\,meV,\cite{widmann:2016b} not too far from the values determined from these optical experiments. In the case of the Ge compound, the strong blue shift saturates below 100\,K. Again, both anomalies, the structural as well as the magnetic phase transition can easily be identified in the temperature dependence of the gap energies. The energy gap seems to be slightly suppressed in the orbitally ordered but paramagnetic phase.

\begin{figure}[b]
\centering
\includegraphics[width=\linewidth]{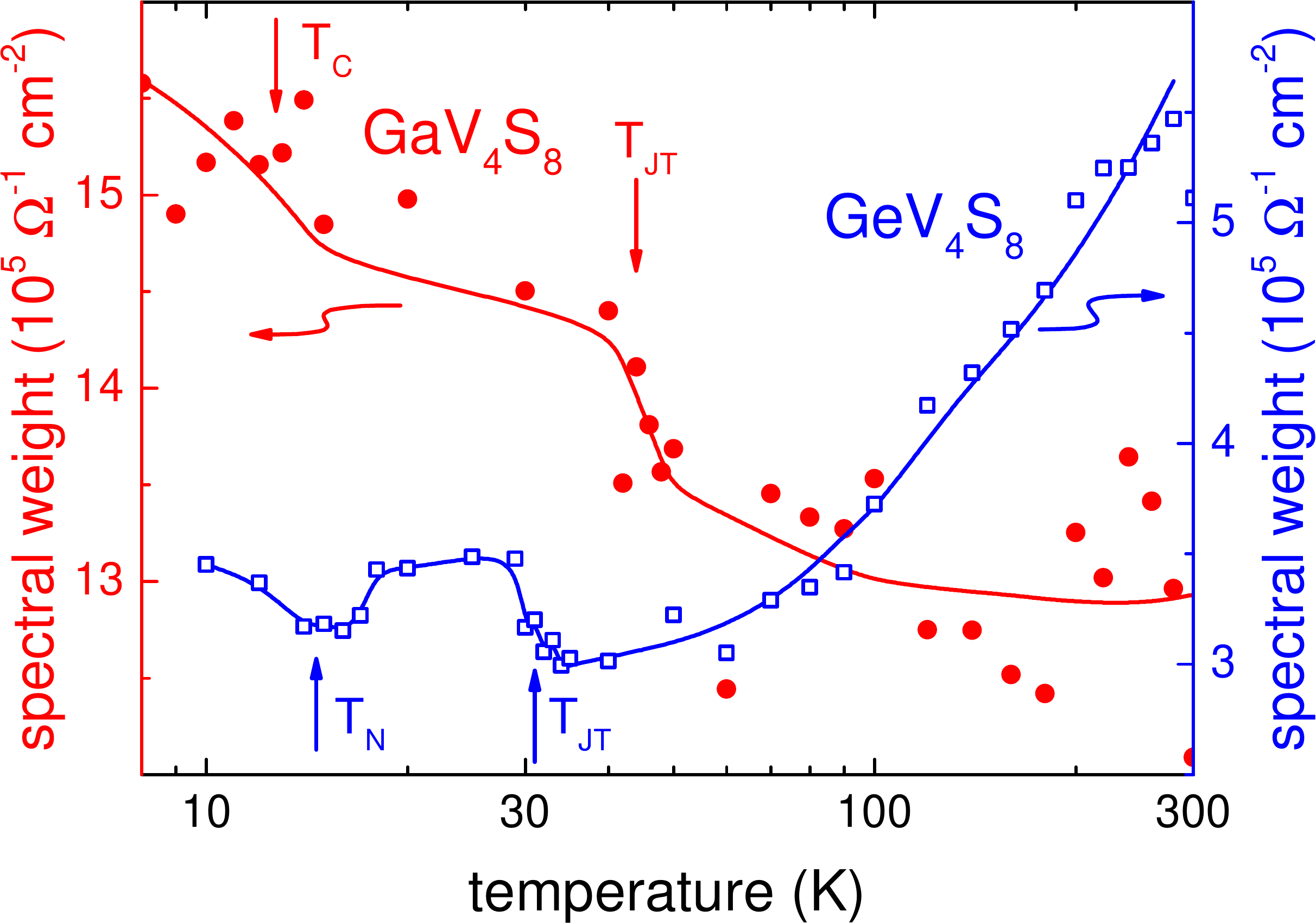}
\caption{Temperature dependence of the spectral weight in GaV$_4$S$_8$ (full circles; left scale) and GeV$_4$S$_8$ (open squares; right scale) as determined by integrating the optical conductivities up to 5\,000\,cm$^{-1}$. Arrows indicate the structural and magnetic phase transitions. The solid lines are drawn to guide the eye.}
\label{fig:10}
\end{figure} 

Figures~\ref{fig:6} and ~\ref{fig:7} document an extreme smearing out of the band edges. At low temperatures both lacunar spinels under consideration exhibit well-defined band edges at low temperatures and a drastic soft-edge behavior at room temperature. These smearing-out effects are closely linked with transfer of optical weight and with the occurrence of isosbestic points. Hence, to unravel the nature of this band-edge softening it seems important to estimate the temperature dependence of the optical weight for a given spectral range for both compounds.
The optical or spectral weight is rigorously defined as the area under the conductivity spectrum and only depends on the electronic density or, naively, on the number of electrons. Integrating up to the highest frequencies it must be constant for a given material. Figure~\ref{fig:10} shows the spectral weight for both compounds, accounting for the optical conductivities up to wave numbers of 5\,000\,cm$^{-1}$. 
This upper limit is chosen rather arbitrarily, but in both compounds the estimated wave number regime covers the main smearing effects of the band edges and also covers both isosbestic points. We have to admit that the experimental uncertainties are drastically different in the two compounds. The scatter of the data is large in the Ga compound. In contrast, the temperature dependence of the optical weight is well defined in GeV$_4$S$_8$. Nevertheless, a number of important conclusions can be drawn from this figure. As can be already detected in the raw data presented in Figs.~\ref{fig:6} and \ref{fig:7}, the transfer of optical weight behaves rather different in the two compounds. For the Ga compound, optical weight is steadily decreasing with increasing temperature. There are indications of step-like drops at the magnetic and at the structural phase transition, but these steps are almost within experimental uncertainty. Overall and beyond experimental uncertainty, with increasing temperature optical weight is shifted from low to high energies. We would like to add, that the optical weight in GaV$_4$S$_8$ calculated up to 10\,000 and 15\,000\,cm$^{-1}$ reveals a very similar temperature dependence. We conclude that in this case transfer of optical weight must be to much higher energies beyond 15000\,cm$^{-1}$ ($\approx 2$~eV).

The temperature dependence of the optical weight is completely different for the Ge compound. At low temperatures, the spectral weight is a factor of four lower compared to the GaV$_4$S$_8$. On increasing temperatures, it increases at $T_\mathrm{N}$ and subsequently decreases at $T_\mathrm{JT}$ and then steadily increases up to room temperature gaining almost a factor of two. In contrast to the Ga compound, with increasing temperatures spectral weight is transferred from high to low energies. It is interesting to note that in the Ge compound, when the optical weight is integrated up to 10\,000\,cm$^{-1}$, the spectral weight is constant and temperature independent. This fact documents that in the latter compound optical weight is shifted only at lower wave numbers. To summarize, in GaV$_4$S$_8$ a transfer of optical weight from low to high energies takes place on increasing temperatures. In this compound, the transfer of optical weight obviously involves wave numbers significantly larger than 15\,000\,cm$^{-1}$. In GeV$_4$S$_8$ this transfer is observed at low temperatures, involving wave numbers below 10\,000\,cm$^{-1}$ only. At the moment we have no reasonable explanation for the transfer of optical weight in these two compounds and we have no explanations for the drastically different behavior. Significant band-edge shifts have been observed for a number of spinel compounds\cite{harbeke:1966,lehmann:1970} and in europium chalcogenides\cite{argyle:1965,busch:1966} and have been explained by magneto-elastic couplings.\cite{callen:1968}

\section{Summary and Conclusions}

We have presented a detailed study of the optical reflectivity in the lacunar spinels GaV$_4$S$_8$ and GeV$_4$S$_8$. Phonons and low-lying electronic transitions are studied as function of temperature. The lacunar spinels of this study are cluster compounds with V$_4$ molecules with a unique electronic distribution and well-defined spin. Both compounds show orbital order induced ferroelectricity and complex magnetic phases at low temperatures. Different Jahn-Teller distortions result in different crystal symmetries of the orbitally ordered phases and in cycloidal spin order in the Ga and in antiferromagnetic order in the Ge compound.  

In both compounds, we found four phonon modes in the high-temperature cubic phase, with similar eigenfrequencies but significantly different dielectric strengths. In the temperature dependence of the phonon frequencies and damping coefficients, we detected significant deviations from the canonical behavior of anharmonic solids. At the Jahn-Teller transitions, the splitting of phonon modes is different in the two compounds due to differences in the symmetry of the orbitally ordered phases. No further splitting of modes is observed for the Ga compound in passing to the cycloidal spin order at low temperatures. In marked contrast, the Ge compound, which undergoes antiferromagnetic spin ordering, reveals significant splitting at the magnetic phase boundary. These findings seem to indicate that, in contrast to GaV$_4$S$_8$, the magnetic transition in GeV$_4$S$_8$ is accompanied by lattice distortions and that spin and lattice degrees of freedom are strongly coupled in this compound. This notion is well consistent with measurements of the dielectric constant $\epsilon'$ in both materials: While in GaV$_4$S$_8$, $\epsilon'(T)$ does not exhibit any significant anomalies at the magnetic transition,\cite{widmann:2016a, ruff:2015} in the Ge compound both, the absolute values and the dispersion effects of $\epsilon'$, become strongly suppressed below the antiferromagnetic phase transition at  $T_\mathrm{N}$ (Refs.~\onlinecite{widmann:2016b}, \onlinecite{singh:2014}). This indicates that polar lattice distortions are strongly involved in this transition, which is also in accord with the findings of thermal-expansion measurements performed for GeV$_4$S$_8$ (Ref.~\onlinecite{widmann:2016b}).

The lowest electronic transition in the Ga and in the Ge compound appears close to 3\,000 and 4\,000\,cm$^{-1}$, respectively. Both compounds reveal a smearing out of the band edge on increasing temperatures. We think that this behavior mainly arises from orbital fluctuations when approaching the structural Jahn-Teller transition. At the present stage, it seems unclear if a smearing out of electronic transitions can appear while the phonon modes seem to be much less affected. Of course, as documented in Figs.~\ref{fig:4} and \ref{fig:5}, some modes reveal a rather high damping in the high-temperature cubic phase, but overall all phonon excitations are rather well defined at all temperatures. It is clear that the relevant electronic transitions as documented in Figs.~\ref{fig:6} and \ref{fig:7} represent transitions between the $d$-derived bands of the vanadium ions, which directly participate in the orbital dynamics. However, it also could play a role that orbital fluctuations appear on a time scale faster than vibrational but slower than electronic frequencies. From a linear extrapolation of the optical conductivities, we arrive at an energy gap of 260\,meV for the Ga and of 360\,meV for the Ge compound at room temperature. These values are in reasonable agreement with band gaps as determined from the temperature dependence of the electrical resistance.\cite{widmann:2016a, widmann:2016b} In both compounds we found considerable shift of spectral weight. The shift of spectral weight on decreasing temperature is completely different and exactly opposite for the two compounds under consideration.

\begin{acknowledgments}
We acknowledge partial support by the Deutsche Forschungsgemeinschaft (DFG) via the Transregional Collaborative Research Center TRR~80 ``From Electronic Correlations to Functionality'' (Augsburg, Munich, Stuttgart). This work was supported by Hungarian Research Funds OTKA K 108918, OTKA PD 111756, Bolyai 00565/14/11, by the Lend\"{u}let Program of the Hungarian Academy of Sciences.
\end{acknowledgments}


\begin{thebibliography}{38}%
\makeatletter
\providecommand \@ifxundefined [1]{%
 \@ifx{#1\undefined}
}%
\providecommand \@ifnum [1]{%
 \ifnum #1\expandafter \@firstoftwo
 \else \expandafter \@secondoftwo
 \fi
}%
\providecommand \@ifx [1]{%
 \ifx #1\expandafter \@firstoftwo
 \else \expandafter \@secondoftwo
 \fi
}%
\providecommand \natexlab [1]{#1}%
\providecommand \enquote  [1]{``#1''}%
\providecommand \bibnamefont  [1]{#1}%
\providecommand \bibfnamefont [1]{#1}%
\providecommand \citenamefont [1]{#1}%
\providecommand \href@noop [0]{\@secondoftwo}%
\providecommand \href [0]{\begingroup \@sanitize@url \@href}%
\providecommand \@href[1]{\@@startlink{#1}\@@href}%
\providecommand \@@href[1]{\endgroup#1\@@endlink}%
\providecommand \@sanitize@url [0]{\catcode `\\12\catcode `\$12\catcode
  `\&12\catcode `\#12\catcode `\^12\catcode `\_12\catcode `\%12\relax}%
\providecommand \@@startlink[1]{}%
\providecommand \@@endlink[0]{}%
\providecommand \url  [0]{\begingroup\@sanitize@url \@url }%
\providecommand \@url [1]{\endgroup\@href {#1}{\urlprefix }}%
\providecommand \urlprefix  [0]{URL }%
\providecommand \Eprint [0]{\href }%
\providecommand \doibase [0]{http://dx.doi.org/}%
\providecommand \selectlanguage [0]{\@gobble}%
\providecommand \bibinfo  [0]{\@secondoftwo}%
\providecommand \bibfield  [0]{\@secondoftwo}%
\providecommand \translation [1]{[#1]}%
\providecommand \BibitemOpen [0]{}%
\providecommand \bibitemStop [0]{}%
\providecommand \bibitemNoStop [0]{.\EOS\space}%
\providecommand \EOS [0]{\spacefactor3000\relax}%
\providecommand \BibitemShut  [1]{\csname bibitem#1\endcsname}%
\let\auto@bib@innerbib\@empty
%</preamble>
\bibitem [{\citenamefont {M\"{u}ller}\ \emph {et~al.}(2006)\citenamefont
  {M\"{u}ller}, \citenamefont {Kockelmann},\ and\ \citenamefont
  {Johrendt}}]{mueller:2006}%
  \BibitemOpen
  \bibfield  {author} {\bibinfo {author} {\bibfnamefont {H.}~\bibnamefont
  {M\"{u}ller}}, \bibinfo {author} {\bibfnamefont {W.}~\bibnamefont
  {Kockelmann}}, \ and\ \bibinfo {author} {\bibfnamefont {D.}~\bibnamefont
  {Johrendt}},\ }\href {\doibase 10.1021/cm052809m} {\bibfield  {journal}
  {\bibinfo  {journal} {Chem. Mater.}\ }\textbf {\bibinfo {volume} {18}},\
  \bibinfo {pages} {2174} (\bibinfo {year} {2006})}\BibitemShut {NoStop}%
\bibitem [{\citenamefont {Pocha}\ \emph {et~al.}(2000)\citenamefont {Pocha},
  \citenamefont {Johrendt},\ and\ \citenamefont {P\"{o}ttgen}}]{pocha:2000}%
  \BibitemOpen
  \bibfield  {author} {\bibinfo {author} {\bibfnamefont {R.}~\bibnamefont
  {Pocha}}, \bibinfo {author} {\bibfnamefont {D.}~\bibnamefont {Johrendt}}, \
  and\ \bibinfo {author} {\bibfnamefont {R.}~\bibnamefont {P\"{o}ttgen}},\
  }\href {\doibase 10.1021/cm001099b} {\bibfield  {journal} {\bibinfo
  {journal} {Chem. Mater.}\ }\textbf {\bibinfo {volume} {12}},\ \bibinfo
  {pages} {2882} (\bibinfo {year} {2000})}\BibitemShut {NoStop}%
\bibitem [{\citenamefont {Bichler}\ \emph {et~al.}(2008)\citenamefont
  {Bichler}, \citenamefont {Zinth}, \citenamefont {Johrendt}, \citenamefont
  {Heyer}, \citenamefont {Forthaus}, \citenamefont {Lorenz},\ and\
  \citenamefont {Abd-Elmeguid}}]{bichler:2008}%
  \BibitemOpen
  \bibfield  {author} {\bibinfo {author} {\bibfnamefont {D.}~\bibnamefont
  {Bichler}}, \bibinfo {author} {\bibfnamefont {V.}~\bibnamefont {Zinth}},
  \bibinfo {author} {\bibfnamefont {D.}~\bibnamefont {Johrendt}}, \bibinfo
  {author} {\bibfnamefont {O.}~\bibnamefont {Heyer}}, \bibinfo {author}
  {\bibfnamefont {M.~K.}\ \bibnamefont {Forthaus}}, \bibinfo {author}
  {\bibfnamefont {T.}~\bibnamefont {Lorenz}}, \ and\ \bibinfo {author}
  {\bibfnamefont {M.~M.}\ \bibnamefont {Abd-Elmeguid}},\ }\href {\doibase
  10.1103/PhysRevB.77.212102} {\bibfield  {journal} {\bibinfo  {journal} {Phys.
  Rev. B}\ }\textbf {\bibinfo {volume} {77}},\ \bibinfo {pages} {212102}
  (\bibinfo {year} {2008})}\BibitemShut {NoStop}%
\bibitem [{\citenamefont {Chudo}\ \emph {et~al.}(2006)\citenamefont {Chudo},
  \citenamefont {Michioka}, \citenamefont {Nakamura},\ and\ \citenamefont
  {Yoshimura}}]{chudo:2006}%
  \BibitemOpen
  \bibfield  {author} {\bibinfo {author} {\bibfnamefont {H.}~\bibnamefont
  {Chudo}}, \bibinfo {author} {\bibfnamefont {C.}~\bibnamefont {Michioka}},
  \bibinfo {author} {\bibfnamefont {H.}~\bibnamefont {Nakamura}}, \ and\
  \bibinfo {author} {\bibfnamefont {K.}~\bibnamefont {Yoshimura}},\ }\href
  {\doibase http://dx.doi.org/10.1016/j.physb.2006.01.461} {\bibfield
  {journal} {\bibinfo  {journal} {Physica B}\ }\textbf {\bibinfo {volume}
  {378–-380}},\ \bibinfo {pages} {1150} (\bibinfo {year} {2006})}\BibitemShut
  {NoStop}%
\bibitem [{\citenamefont {Widmann}\ \emph
  {et~al.}(2016{\natexlab{a}})\citenamefont {Widmann}, \citenamefont {Ruff},
  \citenamefont {G\"{u}nther}, \citenamefont {von Nidda}, \citenamefont
  {Lunkenheimer}, \citenamefont {Tsurkan}, \citenamefont {Bord\'{a}cs},
  \citenamefont {K\'{e}zsm\'{a}rki},\ and\ \citenamefont
  {Loidl}}]{widmann:2016a}%
  \BibitemOpen
  \bibfield  {author} {\bibinfo {author} {\bibfnamefont {S.}~\bibnamefont
  {Widmann}}, \bibinfo {author} {\bibfnamefont {E.}~\bibnamefont {Ruff}},
  \bibinfo {author} {\bibfnamefont {A.}~\bibnamefont {G\"{u}nther}}, \bibinfo
  {author} {\bibfnamefont {H.-A.~K.}\ \bibnamefont {von Nidda}}, \bibinfo
  {author} {\bibfnamefont {P.}~\bibnamefont {Lunkenheimer}}, \bibinfo {author}
  {\bibfnamefont {V.}~\bibnamefont {Tsurkan}}, \bibinfo {author} {\bibfnamefont
  {S.}~\bibnamefont {Bord\'{a}cs}}, \bibinfo {author} {\bibfnamefont
  {I.}~\bibnamefont {K\'{e}zsm\'{a}rki}}, \ and\ \bibinfo {author}
  {\bibfnamefont {A.}~\bibnamefont {Loidl}},\ }\href {\doibase
  10.1080/14786435.2016.1253885} {\bibfield  {journal} {\bibinfo  {journal}
  {Philos. Mag.}\ } (\bibinfo {year} {2016}{\natexlab{a}}),\
  doi: 10.1080/14786435.2016.1253885}\BibitemShut {NoStop}%
\bibitem [{\citenamefont {Widmann}\ \emph
  {et~al.}(2016{\natexlab{b}})\citenamefont {Widmann}, \citenamefont
  {G\"{u}nther}, \citenamefont {Ruff}, \citenamefont {Tsurkan}, \citenamefont
  {Krug~von Nidda}, \citenamefont {Lunkenheimer},\ and\ \citenamefont
  {Loidl}}]{widmann:2016b}%
  \BibitemOpen
  \bibfield  {author} {\bibinfo {author} {\bibfnamefont {S.}~\bibnamefont
  {Widmann}}, \bibinfo {author} {\bibfnamefont {A.}~\bibnamefont
  {G\"{u}nther}}, \bibinfo {author} {\bibfnamefont {E.}~\bibnamefont {Ruff}},
  \bibinfo {author} {\bibfnamefont {V.}~\bibnamefont {Tsurkan}}, \bibinfo
  {author} {\bibfnamefont {H.-A.}\ \bibnamefont {Krug~von Nidda}}, \bibinfo
  {author} {\bibfnamefont {P.}~\bibnamefont {Lunkenheimer}}, \ and\ \bibinfo
  {author} {\bibfnamefont {A.}~\bibnamefont {Loidl}},\ }\href {\doibase
  10.1103/PhysRevB.94.214421} {\bibfield  {journal} {\bibinfo  {journal} {Phys.
  Rev. B}\ }\textbf {\bibinfo {volume} {94}},\ \bibinfo {pages} {214421}
  (\bibinfo {year} {2016}{\natexlab{b}})}\BibitemShut {NoStop}%
\bibitem [{\citenamefont {Singh}\ \emph {et~al.}(2014)\citenamefont {Singh},
  \citenamefont {Simon}, \citenamefont {Cannuccia}, \citenamefont {Lepetit},
  \citenamefont {Corraze}, \citenamefont {Janod},\ and\ \citenamefont
  {Cario}}]{singh:2014}%
  \BibitemOpen
  \bibfield  {author} {\bibinfo {author} {\bibfnamefont {K.}~\bibnamefont
  {Singh}}, \bibinfo {author} {\bibfnamefont {C.}~\bibnamefont {Simon}},
  \bibinfo {author} {\bibfnamefont {E.}~\bibnamefont {Cannuccia}}, \bibinfo
  {author} {\bibfnamefont {M.-B.}\ \bibnamefont {Lepetit}}, \bibinfo {author}
  {\bibfnamefont {B.}~\bibnamefont {Corraze}}, \bibinfo {author} {\bibfnamefont
  {E.}~\bibnamefont {Janod}}, \ and\ \bibinfo {author} {\bibfnamefont
  {L.}~\bibnamefont {Cario}},\ }\href {\doibase 10.1103/PhysRevLett.113.137602}
  {\bibfield  {journal} {\bibinfo  {journal} {Phys. Rev. Lett.}\ }\textbf
  {\bibinfo {volume} {113}},\ \bibinfo {pages} {137602} (\bibinfo {year}
  {2014})}\BibitemShut {NoStop}%
\bibitem [{\citenamefont {Ruff}\ \emph {et~al.}(2015)\citenamefont {Ruff},
  \citenamefont {Widmann}, \citenamefont {Lunkenheimer}, \citenamefont
  {Tsurkan}, \citenamefont {Bord\'{a}cs}, \citenamefont {K\'{e}zsm\'{a}rki},\
  and\ \citenamefont {Loidl}}]{ruff:2015}%
  \BibitemOpen
  \bibfield  {author} {\bibinfo {author} {\bibfnamefont {E.}~\bibnamefont
  {Ruff}}, \bibinfo {author} {\bibfnamefont {S.}~\bibnamefont {Widmann}},
  \bibinfo {author} {\bibfnamefont {P.}~\bibnamefont {Lunkenheimer}}, \bibinfo
  {author} {\bibfnamefont {V.}~\bibnamefont {Tsurkan}}, \bibinfo {author}
  {\bibfnamefont {S.}~\bibnamefont {Bord\'{a}cs}}, \bibinfo {author}
  {\bibfnamefont {I.}~\bibnamefont {K\'{e}zsm\'{a}rki}}, \ and\ \bibinfo
  {author} {\bibfnamefont {A.}~\bibnamefont {Loidl}},\ }\href {\doibase
  10.1126/sciadv.1500916} {\bibfield  {journal} {\bibinfo  {journal} {Sci.
  Adv.}\ }\textbf {\bibinfo {volume} {1}},\ \bibinfo {pages} {e1500916}
  (\bibinfo {year} {2015})}\BibitemShut {NoStop}%
\bibitem [{\citenamefont {K\'{e}zsm\'{a}rki}\ \emph {et~al.}(2015)\citenamefont
  {K\'{e}zsm\'{a}rki}, \citenamefont {Bord\'{a}cs}, \citenamefont {Milde},
  \citenamefont {Neuber}, \citenamefont {Eng}, \citenamefont {White},
  \citenamefont {R\o{}nnow}, \citenamefont {Dewhurst}, \citenamefont
  {Mochizuki}, \citenamefont {Yanai}, \citenamefont {Nakamura}, \citenamefont
  {Ehlers}, \citenamefont {Tsurkan},\ and\ \citenamefont
  {Loidl}}]{kezsmarki:2015}%
  \BibitemOpen
  \bibfield  {author} {\bibinfo {author} {\bibfnamefont {I.}~\bibnamefont
  {K\'{e}zsm\'{a}rki}}, \bibinfo {author} {\bibfnamefont {S.}~\bibnamefont
  {Bord\'{a}cs}}, \bibinfo {author} {\bibfnamefont {P.}~\bibnamefont {Milde}},
  \bibinfo {author} {\bibfnamefont {E.}~\bibnamefont {Neuber}}, \bibinfo
  {author} {\bibfnamefont {L.~M.}\ \bibnamefont {Eng}}, \bibinfo {author}
  {\bibfnamefont {J.~S.}\ \bibnamefont {White}}, \bibinfo {author}
  {\bibfnamefont {H.~M.}\ \bibnamefont {R\o{}nnow}}, \bibinfo {author}
  {\bibfnamefont {C.~D.}\ \bibnamefont {Dewhurst}}, \bibinfo {author}
  {\bibfnamefont {M.}~\bibnamefont {Mochizuki}}, \bibinfo {author}
  {\bibfnamefont {K.}~\bibnamefont {Yanai}}, \bibinfo {author} {\bibfnamefont
  {H.}~\bibnamefont {Nakamura}}, \bibinfo {author} {\bibfnamefont
  {D.}~\bibnamefont {Ehlers}}, \bibinfo {author} {\bibfnamefont
  {V.}~\bibnamefont {Tsurkan}}, \ and\ \bibinfo {author} {\bibfnamefont
  {A.}~\bibnamefont {Loidl}},\ }\href {http://dx.doi.org/10.1038/nmat4402}
  {\bibfield  {journal} {\bibinfo  {journal} {Nat. Mater.}\ }\textbf {\bibinfo
  {volume} {14}},\ \bibinfo {pages} {1116} (\bibinfo {year}
  {2015})}\BibitemShut {NoStop}%
\bibitem [{\citenamefont {Hlinka}\ \emph {et~al.}(2016)\citenamefont {Hlinka},
  \citenamefont {Borodavka}, \citenamefont {Rafalovskyi}, \citenamefont
  {Docekalova}, \citenamefont {Pokorny}, \citenamefont {Gregora}, \citenamefont
  {Tsurkan}, \citenamefont {Nakamura}, \citenamefont {Mayr}, \citenamefont
  {Kuntscher}, \citenamefont {Loidl}, \citenamefont {Bord\'{a}cs},
  \citenamefont {Szaller}, \citenamefont {Lee}, \citenamefont {Lee},\ and\
  \citenamefont {K\'{e}zsm\'{a}rki}}]{hlinka:2016}%
  \BibitemOpen
  \bibfield  {author} {\bibinfo {author} {\bibfnamefont {J.}~\bibnamefont
  {Hlinka}}, \bibinfo {author} {\bibfnamefont {F.}~\bibnamefont {Borodavka}},
  \bibinfo {author} {\bibfnamefont {I.}~\bibnamefont {Rafalovskyi}}, \bibinfo
  {author} {\bibfnamefont {Z.}~\bibnamefont {Docekalova}}, \bibinfo {author}
  {\bibfnamefont {J.}~\bibnamefont {Pokorny}}, \bibinfo {author} {\bibfnamefont
  {I.}~\bibnamefont {Gregora}}, \bibinfo {author} {\bibfnamefont
  {V.}~\bibnamefont {Tsurkan}}, \bibinfo {author} {\bibfnamefont
  {H.}~\bibnamefont {Nakamura}}, \bibinfo {author} {\bibfnamefont
  {F.}~\bibnamefont {Mayr}}, \bibinfo {author} {\bibfnamefont {C.~A.}\
  \bibnamefont {Kuntscher}}, \bibinfo {author} {\bibfnamefont {A.}~\bibnamefont
  {Loidl}}, \bibinfo {author} {\bibfnamefont {S.}~\bibnamefont {Bord\'{a}cs}},
  \bibinfo {author} {\bibfnamefont {D.}~\bibnamefont {Szaller}}, \bibinfo
  {author} {\bibfnamefont {H.-J.}\ \bibnamefont {Lee}}, \bibinfo {author}
  {\bibfnamefont {J.~H.}\ \bibnamefont {Lee}}, \ and\ \bibinfo {author}
  {\bibfnamefont {I.}~\bibnamefont {K\'{e}zsm\'{a}rki}},\ }\href {\doibase
  10.1103/PhysRevB.94.060104} {\bibfield  {journal} {\bibinfo  {journal} {Phys.
  Rev. B}\ }\textbf {\bibinfo {volume} {94}},\ \bibinfo {pages} {060104}
  (\bibinfo {year} {2016})}\BibitemShut {NoStop}%
\bibitem [{\citenamefont {Xu}\ and\ \citenamefont {Xiang}(2015)}]{xu:2015}%
  \BibitemOpen
  \bibfield  {author} {\bibinfo {author} {\bibfnamefont {K.}~\bibnamefont
  {Xu}}\ and\ \bibinfo {author} {\bibfnamefont {H.~J.}\ \bibnamefont {Xiang}},\
  }\href {\doibase 10.1103/PhysRevB.92.121112} {\bibfield  {journal} {\bibinfo
  {journal} {Phys. Rev. B}\ }\textbf {\bibinfo {volume} {92}},\ \bibinfo
  {pages} {121112} (\bibinfo {year} {2015})}\BibitemShut {NoStop}%
\bibitem [{\citenamefont {Wang}\ \emph {et~al.}(2015)\citenamefont {Wang},
  \citenamefont {Ruff}, \citenamefont {Schmidt}, \citenamefont {Tsurkan},
  \citenamefont {K\'{e}zsm\'{a}rki}, \citenamefont {Lunkenheimer},\ and\
  \citenamefont {Loidl}}]{wang:2015}%
  \BibitemOpen
  \bibfield  {author} {\bibinfo {author} {\bibfnamefont {Zhe}~\bibnamefont
  {Wang}}, \bibinfo {author} {\bibfnamefont {E.}~\bibnamefont {Ruff}}, \bibinfo
  {author} {\bibfnamefont {M.}~\bibnamefont {Schmidt}}, \bibinfo {author}
  {\bibfnamefont {V.}~\bibnamefont {Tsurkan}}, \bibinfo {author} {\bibfnamefont
  {I.}~\bibnamefont {K\'{e}zsm\'{a}rki}}, \bibinfo {author} {\bibfnamefont
  {P.}~\bibnamefont {Lunkenheimer}}, \ and\ \bibinfo {author} {\bibfnamefont
  {A.}~\bibnamefont {Loidl}},\ }\href {\doibase 10.1103/PhysRevLett.115.207601}
  {\bibfield  {journal} {\bibinfo  {journal} {Phys. Rev. Lett.}\ }\textbf
  {\bibinfo {volume} {115}},\ \bibinfo {pages} {207601} (\bibinfo {year}
  {2015})}\BibitemShut {NoStop}%
\bibitem [{\citenamefont {Zhang}\ \emph {et~al.}(2017)\citenamefont {Zhang},
  \citenamefont {Wang}, \citenamefont {Yang}, \citenamefont {Xia},
  \citenamefont {Lu},\ and\ \citenamefont {Zhu}}]{zhang:2017}%
  \BibitemOpen
  \bibfield  {author} {\bibinfo {author} {\bibfnamefont {J.~T.}\ \bibnamefont
  {Zhang}}, \bibinfo {author} {\bibfnamefont {J.~L.}\ \bibnamefont {Wang}},
  \bibinfo {author} {\bibfnamefont {X.~Q.}\ \bibnamefont {Yang}}, \bibinfo
  {author} {\bibfnamefont {W.~S.}\ \bibnamefont {Xia}}, \bibinfo {author}
  {\bibfnamefont {X.~M.}\ \bibnamefont {Lu}}, \ and\ \bibinfo {author}
  {\bibfnamefont {J.~S.}\ \bibnamefont {Zhu}},\ }\href {\doibase
  10.1103/PhysRevB.95.085136} {\bibfield  {journal} {\bibinfo  {journal} {Phys.
  Rev. B}\ }\textbf {\bibinfo {volume} {95}},\ \bibinfo {pages} {085136}
  (\bibinfo {year} {2017})}\BibitemShut {NoStop}%
\bibitem [{\citenamefont {Cannuccia}\ \emph {et~al.}(2017)\citenamefont
  {Cannuccia}, \citenamefont {Ta~Phuoc}, \citenamefont {Bri\`{e}re},
  \citenamefont {Cario}, \citenamefont {Janod}, \citenamefont {Corraze},\ and\
  \citenamefont {Lepetit}}]{cannuccia:2017}%
  \BibitemOpen
  \bibfield  {author} {\bibinfo {author} {\bibfnamefont {E.}~\bibnamefont
  {Cannuccia}}, \bibinfo {author} {\bibfnamefont {V.}~\bibnamefont {Ta~Phuoc}},
  \bibinfo {author} {\bibfnamefont {B.}~\bibnamefont {Bri\`{e}re}}, \bibinfo
  {author} {\bibfnamefont {L.}~\bibnamefont {Cario}}, \bibinfo {author}
  {\bibfnamefont {E.}~\bibnamefont {Janod}}, \bibinfo {author} {\bibfnamefont
  {B.}~\bibnamefont {Corraze}}, \ and\ \bibinfo {author} {\bibfnamefont
  {M.~B.}\ \bibnamefont {Lepetit}},\ }\href {\doibase
  http://dx.doi.org/10.1021/acs.jpcc.6b10582} {\bibfield  {journal} {\bibinfo
  {journal} {J. Phys. Chem. C}\ }\textbf {\bibinfo {volume} {121}},\ \bibinfo
  {pages} {3522} (\bibinfo {year} {2017})}\BibitemShut {NoStop}%
\bibitem [{\citenamefont {Warren}\ \emph {et~al.}(2017)\citenamefont {Warren},
  \citenamefont {Pokharel}, \citenamefont {Christianson}, \citenamefont
  {Mandrus},\ and\ \citenamefont {Vald\'es~Aguilar}}]{warren:2017}%
  \BibitemOpen
  \bibfield  {author} {\bibinfo {author} {\bibfnamefont {M.~T.}\ \bibnamefont
  {Warren}}, \bibinfo {author} {\bibfnamefont {G.}~\bibnamefont {Pokharel}},
  \bibinfo {author} {\bibfnamefont {A.~D.}\ \bibnamefont {Christianson}},
  \bibinfo {author} {\bibfnamefont {D.}~\bibnamefont {Mandrus}}, \ and\
  \bibinfo {author} {\bibfnamefont {R.}~\bibnamefont {Vald\'es~Aguilar}},\
  }\href {\doibase 10.1103/PhysRevB.96.054432} {\bibfield  {journal} {\bibinfo
  {journal} {Phys. Rev. B}\ }\textbf {\bibinfo {volume} {96}},\ \bibinfo
  {pages} {054432} (\bibinfo {year} {2017})}\BibitemShut {NoStop}%
\bibitem [{\citenamefont {Reschke}\ \emph {et~al.}(2017)\citenamefont
  {Reschke}, \citenamefont {Wang}, \citenamefont {Mayr}, \citenamefont {Ruff},
  \citenamefont {Lunkenheimer}, \citenamefont {Tsurkan},\ and\ \citenamefont
  {Loidl}}]{reschke:2017b}%
  \BibitemOpen
  \bibfield  {author} {\bibinfo {author} {\bibfnamefont {S.}~\bibnamefont
  {Reschke}}, \bibinfo {author} {\bibfnamefont {Z.}~\bibnamefont {Wang}},
  \bibinfo {author} {\bibfnamefont {F.}~\bibnamefont {Mayr}}, \bibinfo {author}
  {\bibfnamefont {E.}~\bibnamefont {Ruff}}, \bibinfo {author} {\bibfnamefont
  {P.}~\bibnamefont {Lunkenheimer}}, \bibinfo {author} {\bibfnamefont
  {V.}~\bibnamefont {Tsurkan}}, \ and\ \bibinfo {author} {\bibfnamefont
  {A.}~\bibnamefont {Loidl}},\ }\href {https://arxiv.org/pdf/1705.07055}
  {\bibfield  {journal} {\bibinfo  {journal} {arXiv:1705.07055}\ } (\bibinfo
  {year} {2017})}\BibitemShut {NoStop}%
\bibitem [{\citenamefont {Baltensperger}\ and\ \citenamefont
  {Helman}(1968)}]{baltensperger:1968}%
  \BibitemOpen
  \bibfield  {author} {\bibinfo {author} {\bibfnamefont {W.}~\bibnamefont
  {Baltensperger}}\ and\ \bibinfo {author} {\bibfnamefont {J.}~\bibnamefont
  {Helman}},\ }\href
  {http://www.e-periodica.ch/digbib/view?pid=hpa-001:1968:41::678} {\bibfield
  {journal} {\bibinfo  {journal} {Helv. Phys. Acta}\ }\textbf {\bibinfo
  {volume} {41}},\ \bibinfo {pages} {668} (\bibinfo {year} {1968})}\BibitemShut
  {NoStop}%
\bibitem [{\citenamefont {Br\"{u}esch}\ and\ \citenamefont
  {D'Ambrogio}(1972)}]{brueesch:1972}%
  \BibitemOpen
  \bibfield  {author} {\bibinfo {author} {\bibfnamefont {P.}~\bibnamefont
  {Br\"{u}esch}}\ and\ \bibinfo {author} {\bibfnamefont {F.}~\bibnamefont
  {D'Ambrogio}},\ }\href {\doibase 10.1002/pssb.2220500212} {\bibfield
  {journal} {\bibinfo  {journal} {Phys. Status Solidi B}\ }\textbf {\bibinfo
  {volume} {50}},\ \bibinfo {pages} {513} (\bibinfo {year} {1972})}\BibitemShut
  {NoStop}%
\bibitem [{\citenamefont {Sushkov}\ \emph {et~al.}(2005)\citenamefont
  {Sushkov}, \citenamefont {Tchernyshyov}, \citenamefont {Ratcliff~II},
  \citenamefont {Cheong},\ and\ \citenamefont {Drew}}]{sushkov:2005}%
  \BibitemOpen
  \bibfield  {author} {\bibinfo {author} {\bibfnamefont {A.~B.}\ \bibnamefont
  {Sushkov}}, \bibinfo {author} {\bibfnamefont {O.}~\bibnamefont
  {Tchernyshyov}}, \bibinfo {author} {\bibfnamefont {W.}~\bibnamefont
  {Ratcliff~II}}, \bibinfo {author} {\bibfnamefont {S.~W.}\ \bibnamefont
  {Cheong}}, \ and\ \bibinfo {author} {\bibfnamefont {H.~D.}\ \bibnamefont
  {Drew}},\ }\href {\doibase 10.1103/PhysRevLett.94.137202} {\bibfield
  {journal} {\bibinfo  {journal} {Phys. Rev. Lett.}\ }\textbf {\bibinfo
  {volume} {94}},\ \bibinfo {pages} {137202} (\bibinfo {year}
  {2005})}\BibitemShut {NoStop}%
\bibitem [{\citenamefont {Rudolf}\ \emph {et~al.}(2007)\citenamefont {Rudolf},
  \citenamefont {Kant}, \citenamefont {Mayr}, \citenamefont {Hemberger},
  \citenamefont {Tsurkan},\ and\ \citenamefont {Loidl}}]{rudolf:2007b}%
  \BibitemOpen
  \bibfield  {author} {\bibinfo {author} {\bibfnamefont {T.}~\bibnamefont
  {Rudolf}}, \bibinfo {author} {\bibfnamefont {Ch.}~\bibnamefont {Kant}},
  \bibinfo {author} {\bibfnamefont {F.}~\bibnamefont {Mayr}}, \bibinfo {author}
  {\bibfnamefont {J.}~\bibnamefont {Hemberger}}, \bibinfo {author}
  {\bibfnamefont {V.}~\bibnamefont {Tsurkan}}, \ and\ \bibinfo {author}
  {\bibfnamefont {A.}~\bibnamefont {Loidl}},\ }\href
  {http://stacks.iop.org/1367-2630/9/i=3/a=076} {\bibfield  {journal} {\bibinfo
   {journal} {New J. Phys.}\ }\textbf {\bibinfo {volume} {9}},\ \bibinfo
  {pages} {76} (\bibinfo {year} {2007})}\BibitemShut {NoStop}%
\bibitem [{\citenamefont {Harbeke}\ and\ \citenamefont
  {Pinch}(1966)}]{harbeke:1966}%
  \BibitemOpen
  \bibfield  {author} {\bibinfo {author} {\bibfnamefont {G.}~\bibnamefont
  {Harbeke}}\ and\ \bibinfo {author} {\bibfnamefont {H.}~\bibnamefont
  {Pinch}},\ }\href {\doibase 10.1103/PhysRevLett.17.1090} {\bibfield
  {journal} {\bibinfo  {journal} {Phys. Rev. Lett.}\ }\textbf {\bibinfo
  {volume} {17}},\ \bibinfo {pages} {1090} (\bibinfo {year}
  {1966})}\BibitemShut {NoStop}%
\bibitem [{\citenamefont {Lehmann}\ and\ \citenamefont
  {Harbeke}(1970)}]{lehmann:1970}%
  \BibitemOpen
  \bibfield  {author} {\bibinfo {author} {\bibfnamefont {H.~W.}\ \bibnamefont
  {Lehmann}}\ and\ \bibinfo {author} {\bibfnamefont {G.}~\bibnamefont
  {Harbeke}},\ }\href {\doibase 10.1103/PhysRevB.1.319} {\bibfield  {journal}
  {\bibinfo  {journal} {Phys. Rev. B}\ }\textbf {\bibinfo {volume} {1}},\
  \bibinfo {pages} {319} (\bibinfo {year} {1970})}\BibitemShut {NoStop}%
\bibitem [{\citenamefont {Kuzmenko}(2016)}]{reffit:1.2.99}%
  \BibitemOpen
  \bibfield  {author} {\bibinfo {author} {\bibfnamefont {A.}~\bibnamefont
  {Kuzmenko}},\ }\href {https://sites.google.com/site/reffitprogram/home}
  {\enquote {\bibinfo {title} {{RefFIT v. 1.2.99}},}\ }\bibinfo {howpublished}
  {University of Geneva} (\bibinfo {year} {2016}),\ \bibinfo {note}
  {https://sites.google.com/site/reffitprogram/home}\BibitemShut {NoStop}%
\bibitem [{\citenamefont {Kant}\ \emph {et~al.}(2012)\citenamefont {Kant},
  \citenamefont {Schmidt}, \citenamefont {Wang}, \citenamefont {Mayr},
  \citenamefont {Tsurkan}, \citenamefont {Deisenhofer},\ and\ \citenamefont
  {Loidl}}]{kant:2012}%
  \BibitemOpen
  \bibfield  {author} {\bibinfo {author} {\bibfnamefont {Ch.}~\bibnamefont
  {Kant}}, \bibinfo {author} {\bibfnamefont {M.}~\bibnamefont {Schmidt}},
  \bibinfo {author} {\bibfnamefont {Zhe}~\bibnamefont {Wang}}, \bibinfo {author}
  {\bibfnamefont {F.}~\bibnamefont {Mayr}}, \bibinfo {author} {\bibfnamefont
  {V.}~\bibnamefont {Tsurkan}}, \bibinfo {author} {\bibfnamefont
  {J.}~\bibnamefont {Deisenhofer}}, \ and\ \bibinfo {author} {\bibfnamefont
  {A.}~\bibnamefont {Loidl}},\ }\href {\doibase 10.1103/PhysRevLett.108.177203}
  {\bibfield  {journal} {\bibinfo  {journal} {Phys. Rev. Lett.}\ }\textbf
  {\bibinfo {volume} {108}},\ \bibinfo {pages} {177203} (\bibinfo {year}
  {2012})}\BibitemShut {NoStop}%
\bibitem [{\citenamefont {Bord\'{a}cs}\ \emph {et~al.}(2009)\citenamefont
  {Bord\'{a}cs}, \citenamefont {Varjas}, \citenamefont {K\'{e}zsm\'{a}rki},
  \citenamefont {Mih\'{a}ly}, \citenamefont {Baldassarre}, \citenamefont
  {Abouelsayed}, \citenamefont {Kuntscher}, \citenamefont {Ohgushi},\ and\
  \citenamefont {Tokura}}]{bordacs:2009}%
  \BibitemOpen
  \bibfield  {author} {\bibinfo {author} {\bibfnamefont {S.}~\bibnamefont
  {Bord\'{a}cs}}, \bibinfo {author} {\bibfnamefont {D.}~\bibnamefont {Varjas}},
  \bibinfo {author} {\bibfnamefont {I.}~\bibnamefont {K\'{e}zsm\'{a}rki}},
  \bibinfo {author} {\bibfnamefont {G.}~\bibnamefont {Mih\'{a}ly}}, \bibinfo
  {author} {\bibfnamefont {L.}~\bibnamefont {Baldassarre}}, \bibinfo {author}
  {\bibfnamefont {A.}~\bibnamefont {Abouelsayed}}, \bibinfo {author}
  {\bibfnamefont {C.~A.}\ \bibnamefont {Kuntscher}}, \bibinfo {author}
  {\bibfnamefont {K.}~\bibnamefont {Ohgushi}}, \ and\ \bibinfo {author}
  {\bibfnamefont {Y.}~\bibnamefont {Tokura}},\ }\href {\doibase
  10.1103/PhysRevLett.103.077205} {\bibfield  {journal} {\bibinfo  {journal}
  {Phys. Rev. Lett.}\ }\textbf {\bibinfo {volume} {103}},\ \bibinfo {pages}
  {077205} (\bibinfo {year} {2009})}\BibitemShut {NoStop}%
 \bibitem [{\citenamefont {Lee}\ \emph {et~al.}(2004)\citenamefont {Lee},
    \citenamefont {Noh}, \citenamefont {Bae}, \citenamefont {Yang}, \citenamefont
    {Takeda},\ and\ \citenamefont {Kanno}}]{lee:2004}%
    \BibitemOpen
    \bibfield  {author} {\bibinfo {author} {\bibfnamefont {J.~S.}\ \bibnamefont
    {Lee}}, \bibinfo {author} {\bibfnamefont {T.~W.}\ \bibnamefont {Noh}},
    \bibinfo {author} {\bibfnamefont {J.~S.}\ \bibnamefont {Bae}}, \bibinfo
    {author} {\bibfnamefont {I.-S.}\ \bibnamefont {Yang}}, \bibinfo {author}
    {\bibfnamefont {T.}~\bibnamefont {Takeda}}, \ and\ \bibinfo {author}
    {\bibfnamefont {R.}~\bibnamefont {Kanno}},\ }\href {\doibase
    10.1103/PhysRevB.69.214428} {\bibfield  {journal} {\bibinfo  {journal} {Phys.
    Rev. B}\ }\textbf {\bibinfo {volume} {69}},\ \bibinfo {pages} {214428}
    (\bibinfo {year} {2004})}\BibitemShut {NoStop}%
\bibitem [{\citenamefont {Cowley}(1965)}]{cowley:1965}%
  \BibitemOpen
  \bibfield  {author} {\bibinfo {author} {\bibfnamefont {R.~A.}\ \bibnamefont
  {Cowley}},\ }\href {\doibase 10.1051/jphys:019650026011065900} {\bibfield
  {journal} {\bibinfo  {journal} {J. Phys. (Paris)}\ }\textbf {\bibinfo
  {volume} {26}},\ \bibinfo {pages} {659} (\bibinfo {year} {1965})}\BibitemShut
  {NoStop}%
\bibitem [{\citenamefont {Klemens}(1966)}]{klemens:1966}%
  \BibitemOpen
  \bibfield  {author} {\bibinfo {author} {\bibfnamefont {P.~G.}\ \bibnamefont
  {Klemens}},\ }\href {\doibase 10.1103/PhysRev.148.845} {\bibfield  {journal}
  {\bibinfo  {journal} {Phys. Rev.}\ }\textbf {\bibinfo {volume} {148}},\
  \bibinfo {pages} {845} (\bibinfo {year} {1966})}\BibitemShut {NoStop}%
\bibitem [{\citenamefont {Men\'{e}ndez}\ and\ \citenamefont
  {Cardona}(1984)}]{menendez:1984}%
  \BibitemOpen
  \bibfield  {author} {\bibinfo {author} {\bibfnamefont {J.}~\bibnamefont
  {Men\'{e}ndez}}\ and\ \bibinfo {author} {\bibfnamefont {M.}~\bibnamefont
  {Cardona}},\ }\href {\doibase 10.1103/PhysRevB.29.2051} {\bibfield  {journal}
  {\bibinfo  {journal} {Phys. Rev. B}\ }\textbf {\bibinfo {volume} {29}},\
  \bibinfo {pages} {2051} (\bibinfo {year} {1984})}\BibitemShut {NoStop}%
\bibitem [{\citenamefont {Balkanski}\ \emph {et~al.}(1983)\citenamefont
  {Balkanski}, \citenamefont {Wallis},\ and\ \citenamefont
  {Haro}}]{balkanski:1983}%
  \BibitemOpen
  \bibfield  {author} {\bibinfo {author} {\bibfnamefont {M.}~\bibnamefont
  {Balkanski}}, \bibinfo {author} {\bibfnamefont {R.~F.}\ \bibnamefont
  {Wallis}}, \ and\ \bibinfo {author} {\bibfnamefont {E.}~\bibnamefont
  {Haro}},\ }\href {\doibase 10.1103/PhysRevB.28.1928} {\bibfield  {journal}
  {\bibinfo  {journal} {Phys. Rev. B}\ }\textbf {\bibinfo {volume} {28}},\
  \bibinfo {pages} {1928} (\bibinfo {year} {1983})}\BibitemShut {NoStop}%
\bibitem [{\citenamefont {Choi}\ \emph {et~al.}(2003)\citenamefont {Choi},
  \citenamefont {Pashkevich}, \citenamefont {Lamonova}, \citenamefont
  {Kageyama}, \citenamefont {Ueda},\ and\ \citenamefont {Lemmens}}]{choi:2003}%
  \BibitemOpen
  \bibfield  {author} {\bibinfo {author} {\bibfnamefont {K.-Y.}\ \bibnamefont
  {Choi}}, \bibinfo {author} {\bibfnamefont {Y.~G.}\ \bibnamefont
  {Pashkevich}}, \bibinfo {author} {\bibfnamefont {K.~V.}\ \bibnamefont
  {Lamonova}}, \bibinfo {author} {\bibfnamefont {H.}~\bibnamefont {Kageyama}},
  \bibinfo {author} {\bibfnamefont {Y.}~\bibnamefont {Ueda}}, \ and\ \bibinfo
  {author} {\bibfnamefont {P.}~\bibnamefont {Lemmens}},\ }\href {\doibase
  10.1103/PhysRevB.68.104418} {\bibfield  {journal} {\bibinfo  {journal} {Phys.
  Rev. B}\ }\textbf {\bibinfo {volume} {68}},\ \bibinfo {pages} {104418}
  (\bibinfo {year} {2003})}\BibitemShut {NoStop}%
\bibitem [{\citenamefont {Wakamura}\ and\ \citenamefont
  {Arai}(1988)}]{wakamura:1988}%
  \BibitemOpen
  \bibfield  {author} {\bibinfo {author} {\bibfnamefont {K.}~\bibnamefont
  {Wakamura}}\ and\ \bibinfo {author} {\bibfnamefont {T.}~\bibnamefont
  {Arai}},\ }\href {\doibase 10.1063/1.340321} {\bibfield  {journal} {\bibinfo
  {journal} {J. Appl. Phys.}\ }\textbf {\bibinfo {volume} {63}},\ \bibinfo
  {pages} {5824} (\bibinfo {year} {1988})}\BibitemShut {NoStop}%
\bibitem [{\citenamefont {Thomas}\ \emph {et~al.}(1994)\citenamefont {Thomas},
  \citenamefont {Rapkine}, \citenamefont {Carter}, \citenamefont {Millis},
  \citenamefont {Rosenbaum}, \citenamefont {Metcalf},\ and\ \citenamefont
  {Honig}}]{thomas:1994}%
  \BibitemOpen
  \bibfield  {author} {\bibinfo {author} {\bibfnamefont {G.~A.}\ \bibnamefont
  {Thomas}}, \bibinfo {author} {\bibfnamefont {D.~H.}\ \bibnamefont {Rapkine}},
  \bibinfo {author} {\bibfnamefont {S.~A.}\ \bibnamefont {Carter}}, \bibinfo
  {author} {\bibfnamefont {A.~J.}\ \bibnamefont {Millis}}, \bibinfo {author}
  {\bibfnamefont {T.~F.}\ \bibnamefont {Rosenbaum}}, \bibinfo {author}
  {\bibfnamefont {P.}~\bibnamefont {Metcalf}}, \ and\ \bibinfo {author}
  {\bibfnamefont {J.~M.}\ \bibnamefont {Honig}},\ }\href {\doibase
  10.1103/PhysRevLett.73.1529} {\bibfield  {journal} {\bibinfo  {journal}
  {Phys. Rev. Lett.}\ }\textbf {\bibinfo {volume} {73}},\ \bibinfo {pages}
  {1529} (\bibinfo {year} {1994})}\BibitemShut {NoStop}%
\bibitem [{\citenamefont {Greger}\ \emph {et~al.}(2013)\citenamefont {Greger},
  \citenamefont {Kollar},\ and\ \citenamefont {Vollhardt}}]{greger:2013}%
  \BibitemOpen
  \bibfield  {author} {\bibinfo {author} {\bibfnamefont {M.}~\bibnamefont
  {Greger}}, \bibinfo {author} {\bibfnamefont {M.}~\bibnamefont {Kollar}}, \
  and\ \bibinfo {author} {\bibfnamefont {D.}~\bibnamefont {Vollhardt}},\ }\href
  {\doibase 10.1103/PhysRevB.87.195140} {\bibfield  {journal} {\bibinfo
  {journal} {Phys. Rev. B}\ }\textbf {\bibinfo {volume} {87}},\ \bibinfo
  {pages} {195140} (\bibinfo {year} {2013})}\BibitemShut {NoStop}%
\bibitem [{\citenamefont {Wang}\ \emph {et~al.}(2014)\citenamefont {Wang},
  \citenamefont {Schmidt}, \citenamefont {Fischer}, \citenamefont {Tsurkan},
  \citenamefont {Greger}, \citenamefont {Vollhardt}, \citenamefont {Loidl},\
  and\ \citenamefont {Deisenhofer}}]{wang:2014}%
  \BibitemOpen
  \bibfield  {author} {\bibinfo {author} {\bibfnamefont {Zhe}~\bibnamefont
  {Wang}}, \bibinfo {author} {\bibfnamefont {M.}~\bibnamefont {Schmidt}},
  \bibinfo {author} {\bibfnamefont {J.}~\bibnamefont {Fischer}}, \bibinfo
  {author} {\bibfnamefont {V.}~\bibnamefont {Tsurkan}}, \bibinfo {author}
  {\bibfnamefont {M.}~\bibnamefont {Greger}}, \bibinfo {author} {\bibfnamefont
  {D.}~\bibnamefont {Vollhardt}}, \bibinfo {author} {\bibfnamefont
  {A.}~\bibnamefont {Loidl}}, \ and\ \bibinfo {author} {\bibfnamefont
  {J.}~\bibnamefont {Deisenhofer}},\ }\href
  {http://dx.doi.org/10.1038/ncomms4202} {\bibfield  {journal} {\bibinfo
  {journal} {Nat. Commun.}\ }\textbf {\bibinfo {volume} {5}},\ \bibinfo {pages}
  {3202} (\bibinfo {year} {2014})}\BibitemShut {NoStop}%
\bibitem [{\citenamefont {Wang}\ \emph {et~al.}(2016)\citenamefont {Wang},
  \citenamefont {Tsurkan}, \citenamefont {Schmidt}, \citenamefont {Loidl},\
  and\ \citenamefont {Deisenhofer}}]{wang:2016}%
  \BibitemOpen
  \bibfield  {author} {\bibinfo {author} {\bibfnamefont {Zhe}~\bibnamefont
  {Wang}}, \bibinfo {author} {\bibfnamefont {V.}~\bibnamefont {Tsurkan}},
  \bibinfo {author} {\bibfnamefont {M.}~\bibnamefont {Schmidt}}, \bibinfo
  {author} {\bibfnamefont {A.}~\bibnamefont {Loidl}}, \ and\ \bibinfo {author}
  {\bibfnamefont {J.}~\bibnamefont {Deisenhofer}},\ }\href {\doibase
  10.1103/PhysRevB.93.104522} {\bibfield  {journal} {\bibinfo  {journal} {Phys.
  Rev. B}\ }\textbf {\bibinfo {volume} {93}},\ \bibinfo {pages} {104522}
  (\bibinfo {year} {2016})}\BibitemShut {NoStop}%
\bibitem [{\citenamefont {Argyle}\ \emph {et~al.}(1965)\citenamefont {Argyle},
  \citenamefont {Suits},\ and\ \citenamefont {Freiser}}]{argyle:1965}%
  \BibitemOpen
  \bibfield  {author} {\bibinfo {author} {\bibfnamefont {B.~E.}\ \bibnamefont
  {Argyle}}, \bibinfo {author} {\bibfnamefont {J.~C.}\ \bibnamefont {Suits}}, \
  and\ \bibinfo {author} {\bibfnamefont {M.~J.}\ \bibnamefont {Freiser}},\
  }\href {\doibase 10.1103/PhysRevLett.15.822} {\bibfield  {journal} {\bibinfo
  {journal} {Phys. Rev. Lett.}\ }\textbf {\bibinfo {volume} {15}},\ \bibinfo
  {pages} {822} (\bibinfo {year} {1965})}\BibitemShut {NoStop}%
\bibitem [{\citenamefont {Busch}\ and\ \citenamefont
  {Wachter}(1966)}]{busch:1966}%
  \BibitemOpen
  \bibfield  {author} {\bibinfo {author} {\bibfnamefont {G.}~\bibnamefont
  {Busch}}\ and\ \bibinfo {author} {\bibfnamefont {P.}~\bibnamefont
  {Wachter}},\ }\href {\doibase 10.1007/BF02422714} {\bibfield  {journal}
  {\bibinfo  {journal} {Phys. kondens. Mater.}\ }\textbf {\bibinfo {volume}
  {5}},\ \bibinfo {pages} {232} (\bibinfo {year} {1966})}\BibitemShut {NoStop}%
\bibitem [{\citenamefont {Callen}(1968)}]{callen:1968}%
  \BibitemOpen
  \bibfield  {author} {\bibinfo {author} {\bibfnamefont {E.}~\bibnamefont
  {Callen}},\ }\href {\doibase 10.1103/PhysRevLett.20.1045} {\bibfield
  {journal} {\bibinfo  {journal} {Phys. Rev. Lett.}\ }\textbf {\bibinfo
  {volume} {20}},\ \bibinfo {pages} {1045} (\bibinfo {year}
  {1968})}\BibitemShut {NoStop}%
\end{thebibliography}
\end{document}